\documentclass[twocolumn,twoside]{pnas-new}
\usepackage{verbatim}
\usepackage{amsmath}
\templatetype{pnasresearcharticle} 
\usepackage{pdfpages}

\title{Effective routing design for remote entanglement generation on quantum networks}

\author[a,b,1]{Changhao Li}
\author[c,1,2]{Tianyi Li} 
\author[a,b]{Yi-Xiang Liu}
\author[a,b,2]{Paola Cappellaro}

\affil[a]{Research Laboratory of Electronics, Massachusetts Institute of  Technology, Cambridge, MA 02139}
\affil[b]{Department of Nuclear Science and Engineering, Massachusetts Institute of  Technology, Cambridge, MA 02130}
\affil[c]{System Dynamics Group, Sloan School of Management, Massachusetts Institute of  Technology, Cambridge, MA 02139}

\leadauthor{Li} 

\equalauthors{\textsuperscript{1}C. Li and T. Li  contributed equally to this work.}
\correspondingauthor{\textsuperscript{2}Correspondence: tianyil@mit.edu, pcappell@mit.edu}


\begin{abstract}

Quantum network is a promising platform for many ground-breaking applications that lie beyond the capability of its classical counterparts. Efficient entanglement generation on quantum networks with relatively limited resources such as quantum memories is essential to fully realize the network's capabilities, the solution to which calls for delicate network design and is currently at the primitive stage. In this study we propose an effective routing scheme to enable automatic responses for multiple requests of entanglement generation between source-terminal stations on a quantum lattice network with finite edge capacities. Multiple connection paths are exploited for each connection request while entanglement fidelity is ensured for each path by performing entanglement purification. The routing scheme is highly modularized with a flexible nature, embedding quantum operations within the algorithmic workflow, whose performance is evaluated from multiple perspectives. In particular, three algorithms are proposed and compared for the scheduling of capacity allocation on the edges of quantum network. 
Embodying the ideas of \textit{proportional share} and \textit{progressive filling} that have been well-studied in classical routing problems, we design a new scheduling algorithm, the \textit{propagatory update} method, which in certain aspects overrides the two algorithms based on classical heuristics in scheduling performances. The general solution scheme paves the road for effective design of efficient routing and flow control protocols on applicational quantum networks. 
\end{abstract}

\begin{document}

\maketitle
\ifthenelse{\boolean{shortarticle}}{\ifthenelse{\boolean{singlecolumn}}{\abscontentformatted}{\abscontent}}{}

\section*{Introduction}

Quantum networks \cite{Wehnereaam9288} can enable many applications that are beyond the scope of classical data networks, such as quantum communication \cite{Gisin2007NPhon}, clock synchronization \cite{Komar2014NPhy}, and distributed quantum computing \cite{Denchev2008SIGACT,Beals2013PRSA,PhysRevX.4.041041}. Most of these applications require the generation of entangled pairs between far-away stations on the quantum network. 
Recent experiments \cite{DeterEntanglement} have demonstrated  deterministic  entanglement between two spatially-separated solid-state memories via optical photons. The distance in this \textit{physical (hardware) layer} \cite{WehnerSIGCOMM19}  can be further increased with low-loss optical links, such as transporting photons at the tele-communication wavelength \cite{PhysRevLett.123.063601,PhysRevApplied.9.064031,Li_2019}.

\begin{figure}[ht]
\centering
\includegraphics[scale=0.32]{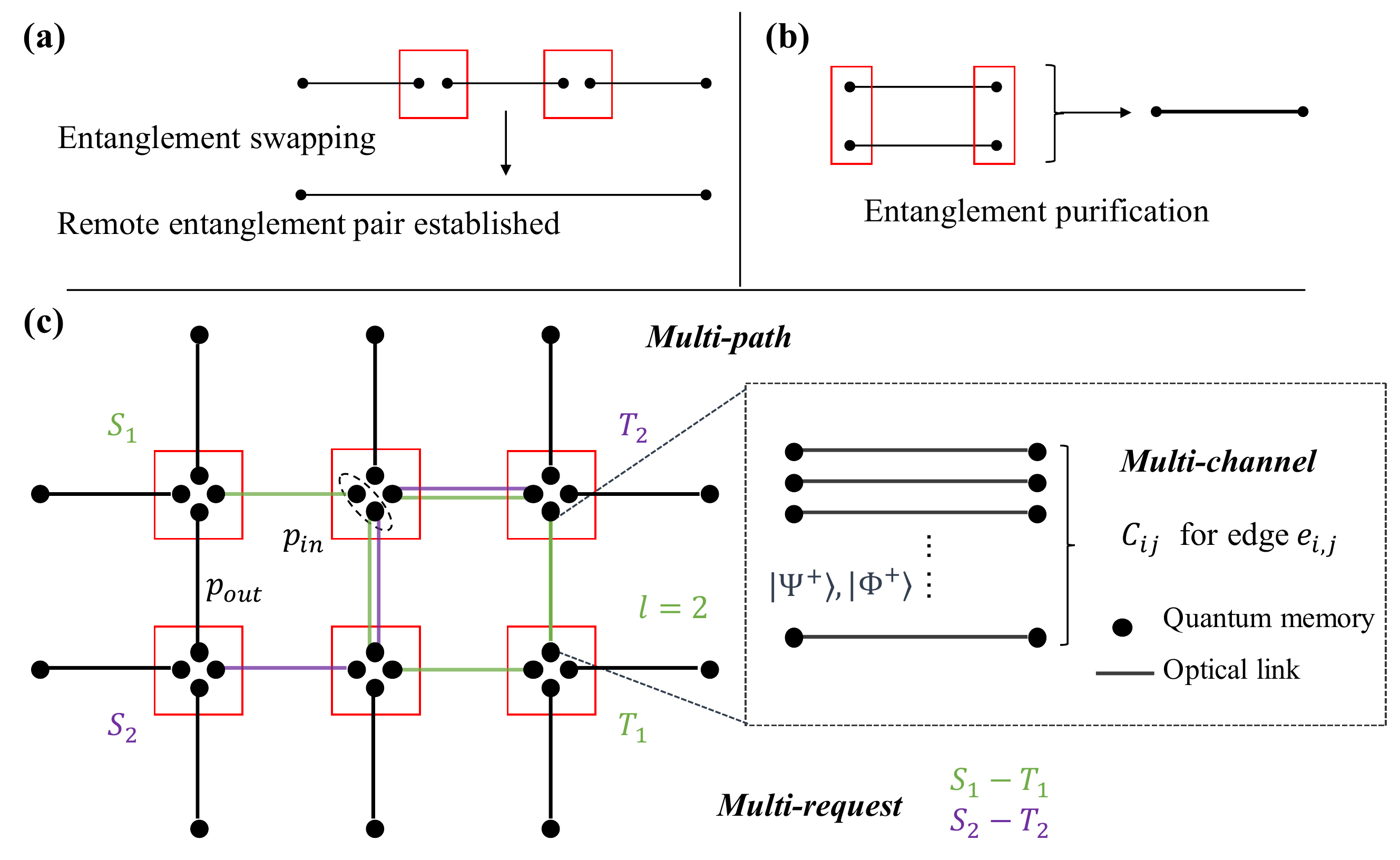}
\caption{\label{Fig1_scheme} Routing problem statement. \textbf{a.} Diagram of entanglement swapping. Two stations that are far away from each other can be entangled with the assistance of intermediate nodes. The black dots (lines) represent quantum memories (optical links).   \textbf{b.} Diagram of entanglement purification. With local operations and classical communication, an entangled pair with high fidelity can be distilled from two fresh pairs with low fidelity.
 \textbf{c.} Entanglement routing in a 2D square lattice network. Edge $(i,j)$ consists of multi entangled pairs (multi-channels) such as Bell states,  and the number of pairs is denoted  capacity $C_{ij}$. Entanglement generation between  $S_{1(2)}$ and $T_{1(2)}$ is requested and there are multi-paths for each requests. The probabilities for successfully building an entangled pair between adjacent nodes and performing local Bell state measurement are denoted as $p_{out}$ and  $p_{in}$, respectively. }
\end{figure}

Still, the time required for remote entanglement generation increases exponentially with distance due to losses. 
In analogy with the \textit{link layer} of classical networks, intermediate nodes serving as quantum repeaters \cite{PhysRevLett.81.5932,DLCZ,PhysRevA.72.052330,RevModPhys.83.33,JLiang2016SR}
could help reduce the exponential time cost in entanglement generation to polynomial time scalings. 
Quantum repeaters (Fig.~\ref{Fig1_scheme}~(a)) work by linking stations over longer distances by performing entanglement swapping, which includes joint Bell state measurements at the local repeater station (see Supplemental Materials) aided by classical communication (LOCC).

Still, quantum repeater might not generate   entangled pairs  with the desired fidelity. Indeed, entangled pairs  is required in many applications, such as quantum cryptography protocols (e.g., the BB84 protocol \cite{BB84}) that require an entanglement fidelity beyond the quantum bit error rate (QBER) to ensure the security of key distribution. 
Then, entanglement purification can increase the fidelity of 
 entangled pairs, at the expense of a reduced number of entangled pairs, i.e., the purification process reduces the number of shared entangled pairs along the links between adjacent nodes on the network (Fig.~\ref{Fig1_scheme}~(b)).  

On top of the link layer mentioned above, where entanglement swapping and purifications are applied,  the \textit{network layer} is required to implement robust and well-functioning network design  to 
enable the broader capabilities of  quantum networks \cite{AAcinNPhy2007,PerseguersNPhy2010,Perseguers_2013,MPant2019npj}. 
In particular, the network design requires routing principles  which enable effective network responses to simultaneous requests of entanglement generation between desired stations (e.g., $S_{1(2)}-T_{1(2)}$ in Fig.~\ref{Fig1_scheme}(c)). Given limited quantum capacity, i.e., a small number of available quantum memories possessed by each node, it is a critical to design an efficient routing protocol for simultaneous entanglement generations between multiple stations.

To tackle this routing problem, we assume that a central processor (scheduler) in the network  conducts routing calculations and then broadcasts the results to all quantum stations, and no local processors on nodes or edges are required. At the beginning of each processing time window, upon receiving a set of connection requests, the processor determines the paths (virtual circuits) to be initiated within the network; then it carries out the scheduling of edge capacity allocation accordingly; and finally it determines the flows of each path according to the scheduling results. Such routing information is broadcast to nodes, and local physical operations, including entanglement purification and swapping, are conducted accordingly, realizing entangled pair connections between remote nodes. 
Entangled pairs between adjacent stations consumed in this time window are generated again to prepare for the next set of  connection requests  to be satisfied in the next time window.

Recent work by M. Pant et.al. \cite{MPant2019npj} introduced a greedy algorithm for this routing problem in a quantum network with square lattice topology. While this algorithm works well on a network with one entangled pair being shared between adjacent nodes, and one connection request being processed per time window, a more general routing protocol where neighboring nodes share more than one entangled pairs (\textit{multi-channel}) and multiple user requests are simultaneously attended to (\textit{multi-request}), still remains an open design question. Novel architectures are to be proposed and tested when this layer of complexity is added to the problem. Moreover, for a robust routing algorithm, it is expected that \textit{multiple paths}, instead of only the shortest path, are to be utilized for each request in order to spread the flow, as a prevention against potential network failures.

In this study, we propose a general routing scheme which can provide efficient solutions to the above entanglement generation problem, with advanced design features (multi-request, multi-path and multi-channel). We design algorithms  to effectively allocate limited quantum resources and meet the needs of entanglement generation for arbitrary connected requests, while satisfying certain fidelity thresholds of generated entanglements.

Routing protocols on classical data networks have been extensively studied and are continuously developed. Yet for the quantum networks  we  consider here, routing schemes well-established on today's communication networks (e.g., the Internet) are not directly applicable. The quantum no-cloning theorem forbids the duplication of entangled pairs and renders unfeasible  the idea of \textit{packet switching}, the  major routing mechanism for large-scale data networks, which sends the same information package multiple times along indefinite routing paths. Instead, for quantum networks, one has to go back to early routing mechanisms that appeared when computers were expensive (although these mechanisms are still in use today for specific applications): the \textit{virtual circuit switching} scheme \cite{Bertsekas1992DataN}, where one establishes temporal paths from two stations and maintain such paths during a time window, and start over again for the next round of connection. Our routing design  is based on this scheme. Conceptually, the key difference between \textit{packet switching} and \textit{virtual circuit switching} is the manner in which the information redundancy is guaranteed, a key issue in practical routing designs in the presence of potential network failures that are frequent and unpredictable in real applications. 
The no-cloning theorem imposes that the redundancy of quantum information cannot be obtained by simply duplicating the entangled pairs, and instead has to be preserved in generating a sufficiently large number of entangled pairs via virtual circuits.

The organization of the paper is as follows: we first formalize the problem statement of this study, based on which we introduce the step-wise routing scheme, in which three algorithms are proposed for scheduling the allocation of adjacent entangled pairs. The performance of the routing results are evaluated from multiple perspectives, after which we discuss the parameters and complexity of the algorithms. A sample routing result is demonstrated and explained in detail. Then we show extensive simulations to discuss system parameters, compare the three scheduling algorithms and test the robustness of the scheme under different conditions. Limitations, scope of use and potential future extensions of the solution scheme are discussed in the end, and the paper is concluded.

\section*{Problem Statement} 
\subsection*{Network parameters and capabilities}
Consider a lattice quantum network $G(V, E)$ of $V$ nodes and $E$ edges. Each node represents a quantum station with a finite number of qubit memories. 
The edge capacity $C_{0}$ is defined as the maximal number of entangled pairs that can be generated between adjacent nodes.  
Without loss of generality, we consider maximally entangled states (Bell states) between adjacent nodes, $|\Phi^{\pm}\rangle= \frac{1}{\sqrt{2}} (|0\rangle|0\rangle \pm |1\rangle|1\rangle)$ and $|\Psi^{\pm}\rangle= \frac{1}{\sqrt{2}} (|0\rangle|1\rangle \pm |1\rangle|0\rangle)$. Since a simple local operation such as a single qubit $X$ or $Z$ gate can transform Bell states from one to another, this provides us the freedom of switching the states at low cost. Different Bell states or states with different entanglement entropy might serve as IDs for entangled pairs for further usages in the network layer \cite{WehnerSIGCOMM19}. 
To ensure entanglement fidelity above a certain threshold $F_{th}$, the network is initialized by quantum entanglement purification on each edge, which results in a reduced number of shared entanglement pairs, thus a reduced edge capacity. 
We assume the fidelities of multiple entangled pairs along the same edge to be identical while the fidelities on different edges can vary, as influenced by various factors such as geographical constraints or human interventions. The average entanglement fidelity of a certain edge can be calculated by pre-characterization; here we assume the fidelities on all edges to follow a normal distribution N($F_{mean}$, $F_{std}$) for simplicity (we take the fidelity to be one if it exceeds unity).

At the start of a time window, a set $R$ of simultaneous connection requests  are received. 
Each request $r\in R$ requires the generation of at least $\bar{f}^r$ entangled pairs between two remote nodes (denoted  source node and terminal node) by implementing entanglement swapping (see Fig.\ref{Fig1_scheme}~(a)) along a given path. 
Local entanglement swapping operations at quantum stations are imperfect, with a probability of success $p_{in}$. This implies that the time needed to successfully entangle two remote stations along a certain path will grow exponentially with increasing path length, an important aspect in evaluating routing performances (see following sections). Optical links between nodes will be successfully established with probability $p_{out}$,   mainly limited by  optical loss during the long distance transmission. We assume homogeneous $p_{in}$ and $p_{out}$ on all network nodes and edges in simulations.

\subsection*{Routing Problem}
With these preliminaries, we define the routing problem on small-capacity regular (lattice) quantum networks as the following: \\
\textit{Given a  quantum network with variable topology due to imperfect initialization and limited quantum memory  on stations, design an  algorithm  that provides routing solutions  to arbitrary entangled-pair requests, able to ensure above-threshold entanglement fidelity and to process multiple connection requests.}

To efficiently utilize network resources within each processing window, multiple connection paths $L^r$ are identified and established for each request $r$, as opposed to only using the single shortest path between the request pair. 
Specifically, the solution scheme determines the flow (allocated capacity) $f_{ij}^{r,l}$ of each path $l$ for each connection request $r$, on every edge $(i,j)$ in the network, such that the aggregated flow of each request $f^{r} = \sum_{l\in L^r}f^{r,l}$ is able to meet the demand $\bar{f}^r$ to a sufficient extent. 

This problem formulation considers multiple paths on the network for a specific connection request, as opposed to only utilizing the (single) shortest path as in most virtual circuit routing schemes \cite{ramakrishnan2001optimal}. We do not aim at a  multi-commodity flow optimization  for routing \cite{frank1971routing}, since the optimization is NP-hard when relaxing the shortest path constraint. 
Although an optimal solution for $f_{ij}^{r,l}$ is not derived here, we strive to provide efficient solutions for this routing task through sophisticated algorithmic design. Note that the queuing process on each node (station) is not considered in this study, which deals with arranging connection requests over sequential processing windows; the current scheme provides routing solutions \textit{within} each processing window, assuming that a certain number of requests are processed in the same batch upon submission.

\section*{Routing Scheme}

Our stepwise routing scheme consists of both physical steps (in which quantum operations are conducted) and algorithmic steps (in which computational planning is performed). The overall scheme, summarizing both computational instructions and physical operations (labeled by red text), is shown below. 

At the start of each processing window (\textit{Step 0}), the quantum network $G = (V,E)$ is reinitialized. 
The maximal edge capacity is $C_0$ for all edges. 
Entanglement generation is attempted between adjacent nodes. Due to the non-unity entanglement generation probability $p_{out}$ between neighbor nodes, the realized capacity is $\leq C_0$.   Multiple transmission requests $R = \{r\}$ are then received. Each request consists of one source and one terminal node ($r = [s,t]$), the  anticipated demand $\bar{f}^r$, as well as a weight $w_r$. There is no restriction on the choice of  source and terminal nodes; the protocol supports all possible scenarios, including same $s$, same $t$, same $[s,t]$. 

\begin{figure*}[ht]
\centering
\includegraphics[scale=0.55]{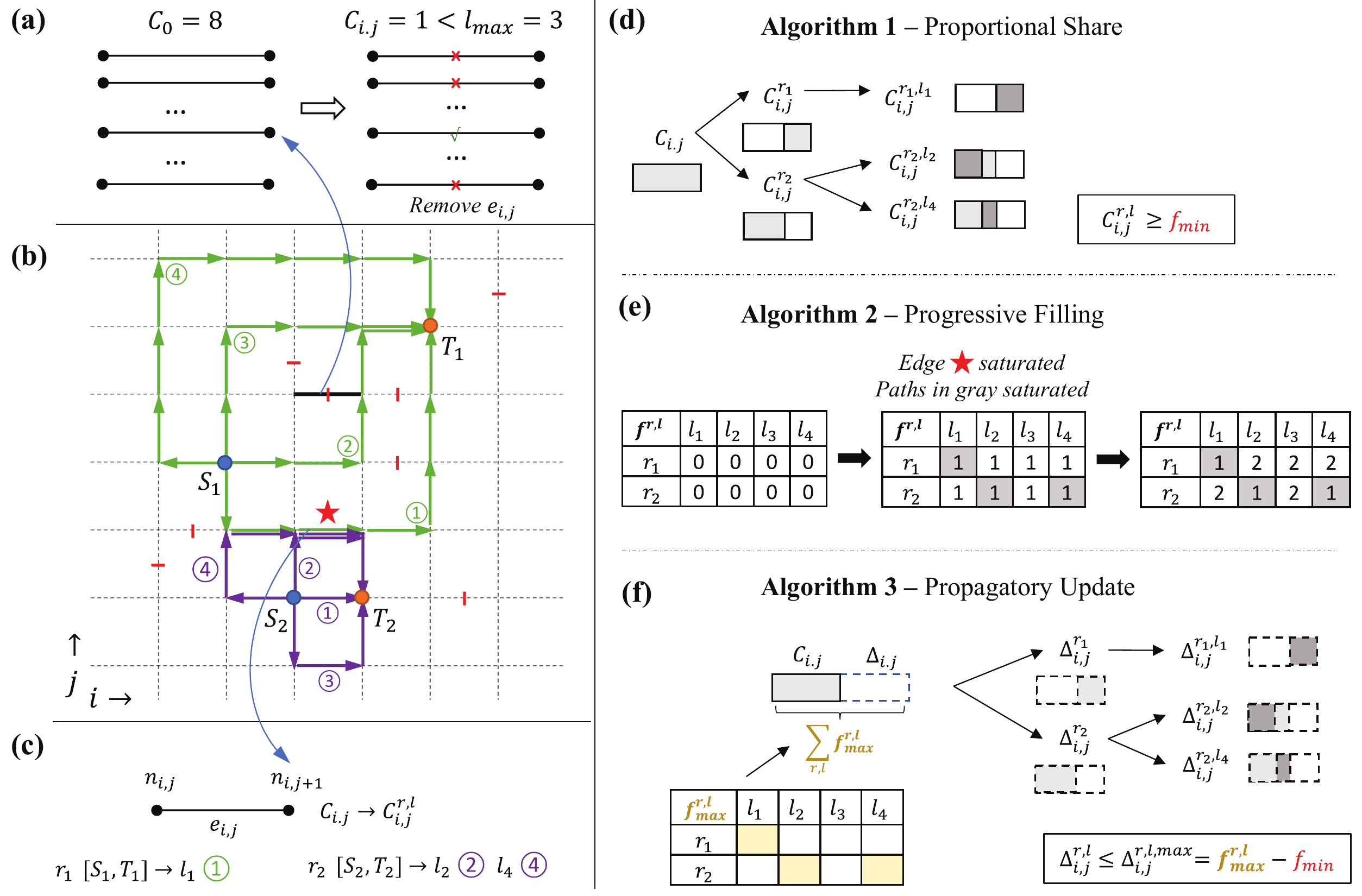}
\caption{\label{Fig2_protocol} Routing scheme \textbf{a.} Initialization of network topology. On one specific edge, purification realizes 1 entanglement pair out of 8 available entangled pairs ($C_{ij} = 1, C_0 = 8$), which falls below the maximum number of paths allowed $l_{max}$. To guarantee that each path has at least one channel allocated, this edge is deactivated (indicated by the red bar). \textbf{b.} Sample routing scenario. Two connection requests ($r_1: S_1\leftrightarrow T_1; r_2: S_2\leftrightarrow T_2$) are received; 4-shortest paths are identified for each request, utilizing active edges (edges without red bars) on the square lattice network. Each edge may be utilized in multiple paths for different connection requests. Here we show the different paths with arrows for simplicity, but we point out that there is no directionality along the paths. \textbf{c.} Capacity allocation scheduling. Three paths for two requests ($l_1$ for $r_1$; $l_2, l_4$ for $r_2$) utilize the starred edge (top left), on which the three algorithms (d-f) are illustrated. 
For \textit{proportional share} and \textit{propagatory update}, entangled pairs on the edge are distributed among the three paths through a two-stage allocation process: the quota is first determined for each request, then for each path of different requests. In \textit{proportional share}, the realized edge capacity is directly allocated;  \textit{propagatory update}  calculates the sum- marized desired capacity for each edge from all paths that utilize it, using the information in the global schedule table, then allocates the unmet capacity demand to other paths. Corresponding deductions from the desired capacities that are beyond the real ones are made, and the schedule table is updated at each step.
The minimum capacity for each path is $f_{min}$; the minimum capacity deduction for path $l$ of request $r$ is $f_{max}^{r,l} - f_{min}$, where $f_{max}^{r,l}$ is the desired capacity in the global routing table. For \textit{progressive filling}, $f^{r,l}$ is 0 on all paths at $t=0$ and increments by 1 at each time step. At $t=1$, the starred edge is saturated and thus the three paths on it are saturated. At $t=2$, $f^{r,l}$ keeps increasing on unsaturated paths.}
\end{figure*}

At \textit{Step 1}, entanglement purification is carried out on the edges $(i,j)\in E$ which have fidelity below the threshold $F_{th}$, and a fraction of the maximum edge capacity $C_{ij} \le  C_0$ is physically realized on each edge. The initial edge fidelity, with normal distribution N($F_{mean}$, $F_{std}$), and the probability of successful entanglements between adjacent stations $p_{out}$ determines the realized capacity. 
We set the fidelity threshold $F_{th}$ that is applied on all edges. The value of $F_{th}$ may vary in practice for different network functionalities. 
To ensure enough bandwidth for each connection path, we apply a cap $l_{max}$ to the number of possible paths traversing a single edge. 
Edges with residual capacity $C_{ij}$ smaller than $l_{max}$ are then deactivated (graphically, removed from $G$), as shown in Fig.~\ref{Fig2_protocol}~(a). 
This ensures that the minimum allocated flow for a single path $f_{min}$, is always greater than one:
\begin{equation}
f_{min} = \textrm{floor}(\frac{\min(C_{ij})}{l_{max}}) \ge \frac{l_{max}}{l_{max}} = 1.
\end{equation}

At \textit{Step 2}, paths for different connection requests are identified in the revised graph $G'$. In order to ensure sufficient redundancy in routing in case of potential network failures,  $k$ shortest paths \citep{Yen1971} are identified between each connection pair $r = [s,t]$. The value of $k$ determines how relaxed the desired paths are in length: when $k=1$ we only consider a single shortest path; more and longer paths will be utilized under a larger $k$. 
A practical and effective value of $k$ should be determined given specific network conditions, such as the number of requests $|R|$ allowed within a time window, the size of the lattice network $|E|$, and the maximum  capacity $C_0$. 
At a minimum, $k$ should be sufficiently large to satisfy at least in principle the  flow demand of each connection request,  $k\, f_{min} \ge \bar{f}^r$.

The identified paths are collected in the  path information set $H=\{h_{ij}\}$. For each edge $(i,j)$ along each path $l$ identified for a request $r$, a new information entry $h_{ij}$ is added to the path information set $H$ with the following form:
\begin{equation}
h_{ij} = [r,l,d,o],
\end{equation}
where $d$ is the path length and $o$ the edge order in $l$.

\textit{Step 3}  is the core procedure of the routing scheme,  setting the  edge-wise capacity allocation scheduling. 
We propose and compare three methods for this task: (i) \textit{proportional share} (PS), (ii) \textit{progressive filling} (PF) and (iii) \textit{propagatory update} (PU), of which (i) and (ii) are based on well-known scheduling heuristics on classical networks and (iii) is a novel algorithm combining the ideas of (i) and (ii). 
Illustrations of the three allocation algorithms are depicted in Fig.~\ref{Fig2_protocol} (d-f).  Here we give only a qualitative description of the algorithms (details are in the Supplemental Materials).

In the \textit{proportional share} method, the scheduling is edge-specific: the capacity $C_{ij}$ is proportionally distributed  to all paths that utilize $(i,j)$, and only local information in $H$ is used. 
The \textit{progressive filling} method implements  the algorithm in \cite{Bertsekas1992DataN}, which should guarantee max-min fairness (as we show in the sections below). 
All paths are treated equally, with flows starting at zero and having uniform incremental increase; since in our setting quantum edge capacities are integers, the increment is simply 1. 
Edges are gradually saturated along the filling process, and paths utilizing saturated edges stop  increasing in flow. 
Our integer scheme follows a slightly different definition of edge saturation than for non-integer flows. 
An edge can be left with unallocated channels that are insufficient to be allocated to the paths going through the edge.
In this case the edge is viewed as saturated with a small leftover capacity. 
The filling process terminates when no new attempt of flow increments is possible; since flows increase in a fair manner, $f_{min}$ is unnecessary for this method. 

Combining ideas from these two conventional methods,  the \textit{propagatory update} method defines a global schedule table and assigns allocation on each edge in a backward, edge-specific manner.
The algorithm calculates  the summarized desired capacity for each edge from all paths that utilize it, using the information in the global schedule table, then allocates the unmet capacity demand to other paths. This requires to make corresponding deductions from the desired capacities that are beyond the real ones, and the schedule table is updated at each step. This is similar to what done in the \textit{progressive filling} method, but there the incremental increases of values in the schedule table are uniform. 
Unlike the \textit{progressive filling}, instead of treating all paths equally, the \textit{proportional share} and the \textit{propagatory update} methods adopt a similar two-stage strategy: the quota is first determined for each request, then for each path of different requests. 
Edge capacities are distributed proportionally at both stages: at the request level, the allocated capacity for each request is proportional to the number of paths of such request that utilize a given edge, to the power $\beta$. 
At the path level, the allocated capacity for each path of a specific request is proportional to the path length, to the power $\alpha$ (see details in Supplemental Materials). 
$\alpha$ and $\beta$ are design features that control the exploitation-exploration tradeoff in path utilization, similar to the idea in the ant colonization algorithm \citep{DB2010}, and essentially determine the fairness of scheduling results. 
They are set as open parameters (not used in the \textit{progressive filling} method) and experimented upon in simulations (see section of results). As mentioned above, before the capacity allocation, the path list $L^r$ going through each edge is truncated to have $l_{max}$ entries, for the PS and PU methods. The criterion for truncation is the path length: keep short paths in the list and remove long paths; a path is always kept in the list if it is the sole path for a certain request. Details of the three scheduling methods are explained in Supplemental Materials.

The final allocated capacity (flow) of each path is determined in \textit{Step 4}. The short-board constraint is applied along each path: for path $l$ of request $r$, the actual flow $f^{r,l}$ is determined by the minimum capacity for this path among all the edges in $l$:
\begin{equation}
f^{r,l} = \min_{l\in L^r}[\ {C^{r,l}_{ij}}],\quad \text{for}\ (i,j)\in l.
\end{equation}
where $C^{r,l}_{ij}$ is the capacity determined in \textit{Step 3}. Note that this constraint is only necessary if \textit{proportional share} scheduling is used at \textit{Step 3}; for both \textit{progressive filling} and \textit{propagatory update} methods, the constraint is already built in the allocation process.

\textit{Step 2-4} complete the algorithmic computation of the routing scheme. Based on the determined routing schedule, local quantum operations (entanglement swapping) are conducted accordingly and scheduled remote entanglements are established (\textit{Step 5}). The actual flow for each request $f^{r} = \sum_{l\in L^r}f^{r,l}$ is compared with the demand $\bar{f}^{r}$; unsatisfied requests are queued to the next processing window and their weight can be adjusted accordingly.

\algdef{SE}[SUBALG]{Indent}{EndIndent}{}{\algorithmicend\ }%
\algtext*{Indent}
\algtext*{EndIndent}
\floatname{algorithm}{}
\begin{algorithm*}[!h]
\caption*{\textbf{Routing Protocol}}
\begin{algorithmic}[1]
\State (Global $F_{mean}$, $F_{std}$, $F_{th}$, $p_{in}$, $p_{out}$).
\State \textbf{Start new processing window}
\State \textbf{Step 0} Initialization
\Indent
   	\State Lattice network $G = (V,E)$ resumed. Initial capacity $C_0$ generated on each edge.
	\State \textcolor{red}{\textbf{Perform entanglement generation between adjacent nodes.}}
	\State Realized capacity less than or equal to $C_0$ on each edge $(i,j)$.
	\State Entanglement generation requests $R = \{r\}$ received with anticipated flow demand $\bar{f}^r$ (and possible weight $W = \{w_r\}$).
\EndIndent
\vspace{10pt}
\State \textbf{Step 1} Entanglement Purification
\Indent
	\State \textcolor{red}{\textbf{Conduct entanglement purification on the edges with fidelity below the required threshold $F_{th}$.}} 
	\State Realize capacity $C_{ij} \le C_0$ on each edge $(i,j)$. 
	\State Determine system parameters $X = \{l_{max}, k, \alpha, \beta\}$ and $f_{min}$. 
	\State Disactivate edges with insufficient capacity $C_{ij} < l_{max}$. $E\rightarrow E'$. $G\rightarrow G'$. 	
\EndIndent
\vspace{10pt}
\State \textbf{Step 2} Path Determination
\Indent
	\For{each $r \in R$}
		\If{No path exists between $S_r$ and $T_r$}
		\State \textbf{Continue}
		\EndIf
		\State Identify $k$ shortest paths $L^r = {\{l\}}_k$ between $S_r$ and $T_r$.
		\For{each $l \in L^r$}
			\State Append tuple $h = [r,l,d,\text{edge\ order}]$ to the path information set $H$ on each edge in $l$.
		\EndFor
	\EndFor
\EndIndent
\vspace{10pt}
\State \textbf{Step 3} Capacity Allocation Scheduling (Algorithm 1-3 in Supplemental Materials)
\Indent
\State Algorithm 1: \textbf{Proportional Share}  
\State Algorithm 2: \textbf{Progressive Filling}  
\State Algorithm 3: \textbf{Propagatory Update}  
\EndIndent
\vspace{10pt}
\State \textbf{Step 4} Flow Determination and Evaluation
\Indent
\State (Proportional Share)
   	\For{each $r\in R$}
		\For{each $l\in L^{r}$}
			\State Determine $f^{r,l} = min\ {C^{r,l}_{ij}}\ \text{for}\ (i,j)\in l$ on path $l$ of request $r$.
		\EndFor
	\EndFor
	\State Calculate $F, u_{ij},\gamma_r, J$. Evaluate routing performance. 
	\State Record scheduling result $\{R,G'\}\Rightarrow \{C_{ij}^{r,l}\}$.
\EndIndent
\vspace{10pt}
\State \textbf{Step 5} Conduct scheduled operations
\Indent
	\State \textcolor{red}{\textbf{Perform entanglement swapping according to the above routing results.}}
	\State Compare realized flow $f^{r} = \sum_{l\in L^r}f^{r,l}$ with demand $\bar{f}^{r}$. Queue unsatisfied demand to the next processing window.
\EndIndent
\end{algorithmic}

\end{algorithm*}

\begin{table*}
\centering
\caption{Evaluation of the routing performance}
\begin{tabular}{lcll}
Property & Notation & Measure explanation & Design feature \\
\midrule
Throughput &  $F$ & Flow: Total number of generated entangled pairs & Multi-request \\
Traffic & $u_{ij}$ & Utilization of capacity on edge $(i,j)$ & Multi-channel  \\
Delay & $\gamma$ & Average path stretching factor & Multi-path \\
\midrule
Fairness over requests & $J_{req}$ & Jain's index: Balance of capacity allocation (for requests)  & Multi-request \\
Fairness over paths & $J_{path}$ & Jain's index: Balance of capacity allocation (for paths)  & Multi-path \\
\end{tabular}
\label{Table:Metrics}
\end{table*}

\section*{Performance Measures}

To evaluate the performance of our routing scheme, we consider three performance measures, targeting the three innovative features of our routing design (\textit{multi-request, multi-channel} and \textit{multi-path}). In line with typical performance evaluation of classical networks, these measures characterize well-established  routing  properties:  \textit{throughput}, \textit{traffic} and \textit{delay}. We also introduce a metric that characterizes the \textit{fairness} of the scheduling scheme.  We summarize these measures in Table. \ref{Table:Metrics}. Since our routing scheme ensures entanglement fidelity above the desired threshold, we do not consider fidelity among the metrics.
  
\textit{Throughput --} Besides fidelity,  entanglement generation rate is the critical figure of merit typically used to evaluate quantum networks. Within a fixed time window, the entanglement generation rate corresponds to the number of successfully established entangled pairs for all  requests. In network routing terminology, this is known as the throughput of the system, a term we adopt in this work. In our setting, we characterize the throughput via the total weighted flow, aggregated over all paths:
\begin{equation}
F = \sum_{r}  w_r \sum_l f^{r,l} p_{in}^{d_{r,l}-1}
\end{equation}

This definition of throughput is slightly different from that for classical networks, due to a non-negligible possibility of failure during internal connections (entanglement swapping) within each station, i.e. $p_{in}\neq 1$. 

Since the routing scheme is not  a multi-commodity flow optimization, and hence it is not guaranteed to satisfy the whole requested demand $\bar{f}^r$, it is plausible in certain cases to ask to maximize   the minimum flow over all requests.
Instead of $F$, then, we could use an alternative metric  to evaluate  the system throughput, the minimum flow among all requests $F_{min} = \min_r w_r \sum_l f^{r,l} p_{in}^{d_{r,l}-1}$.  Tests show that $F$ and $F_{min}$ are almost always positively correlated and thus could be used interchangeably; we thus keep  $F$ as the primary measure for throughput.

\textit{Traffic -- }
To evaluate the routed traffic on the network, we calculate the \textit{utilization factor} $u_{ij}$ of the available capacity  on each edge $(i,j)$:
\begin{equation}
u_{ij} = \frac{\sum_{r,l}f^{r,l}_{ij}}{C_{ij}} \in (0,1 \rbrack.
\end{equation}

Routed traffic could be signaled by the mean $U_{ave}$ and variance $U_{var}$ of the utilization factors $u_{ij}$ on all utilized edges. For an efficient routing performance, it is desired that capacity utilization $U_{ave}$ is high on edges (i.e. little idleness), while the traffic is evenly distributed without many bottleneck edges, i.e. $U_{var}$ is small.

\textit{Delay --}
To preserve sufficient redundancy in routing,  multiple paths of various lengths are exploited for each connection request.  For each request $r$, we calculate the path length stretching factor $\gamma_r$, which is defined as the average length of the request paths (weighted by the flow), normalized by the shortest path length $d_{r,0}$:
\begin{equation}
\gamma_r = \frac{\sum_l f^{r,l}d_{r,l}}{d_{r,0}\sum_l f^{r,l}},
\end{equation}
where $d_{r,l}$ is the path length for a request $r$ and path $l$.
The average stretching factor $\gamma$ of the system is calculated from $\gamma_r$ over all requests; $\gamma_r$ (and thus $\gamma$) is always no less than 1. Quantum connections take nontrivial processing time for each hop on the path, hence longer paths can induce more delay in entanglement generation. If one assumes the same processing time across each edge in the network, then $\gamma$ indicates the average delay of a routing protocol:  for $\gamma = 1$, all flows take the shortest paths and the routing has minimum delay; a large $\gamma$ can correspond to a higher delay where circuitous paths are utilized.

\textit{Fairness --}
Jain's index \cite{jain1984quantitative} is calculated to evaluate the fairness of the scheduling scheme. The total scheduled flow for each request is aggregated over all paths, $f^{r} = \sum_{l\in L^r}f^{r,l}$, and the fairness of the scheduling over requests is given by:
\begin{equation}
J_{req} = \frac{(\sum_{r \in R}w_r  f^{r})^2}{|R|\sum_{r \in R} (w_r f^{r})^2}.
\label{eq:Jreq}
\end{equation}
$J_{req}$ falls between $[0,1]$ with $J_{req} = 1$ representing complete fairness.
Evidently, the \textit{progressive filling} algorithm obtains the highest fairness $(J_{req} \approx 1)$ among the three scheduling methods, guaranteeing max-min fairness \cite{le2005rate}; values of $\alpha$ and $\beta$ in \textit{proportional share} and \textit{propagatory update} algorithms might affect the fairness, which is almost always less than 1. A large $J_{req}$ ensures that connection requests are processed in a relatively fair manner, which is clearly desirable. In a way similar to equation (\ref{eq:Jreq}), one could also calculate the fairness index with respective to each path, $J_{path}$, without aggregating them for the same requests:
\begin{equation}
J_{path} = \frac{(\sum_{r \in R} w_r  \sum_{l\in L^r}f^{r,l})^2}{|R|\sum_{r \in R}w_r^{2}  \sum_{l\in L^r} (f^{r,l})^2}.
\end{equation}
A high fairness of flows over different paths is desirable, in order to avoid a  situation where capacities are concentrated on one of a few paths. The routing will be less robust to potential network failures if one or few paths dominate the flow. 

\section*{Free Parameters Determination}
The routing algorithms rely on a set of free parameters,  $X = \{l_{max}, k, \alpha, \beta\}$ for PS and PU, and $X = k$ for PF. 
As  at the start of each processing window the network is reinitialized, with edge capacities $C_{ij}$ reset according to entanglement purification results, and  new requests $R$ are received, the system parameters need to be determined specifically for the current time window.
The parameters are chosen so as to optimize a desired objective function constructed from the performance measures discussed above. 
The specific form of the objective function can be adapted  to  specific requirements of a given application. 
An exemplary objective function can be obtained by simultaneously considering the throughput $F$, traffic (in terms of $U_{ave}$ and $U_{var}$), and delay $\gamma$, with $\pi_1, \pi_2, \pi_3$ denoting relative weight of each measure.
Then, the corresponding optimization problem  can  be formulated as
\begin{equation}
\begin{split} 
X^* &= \underset{\{l_{max}, k, \alpha, \beta\}}{\mathrm{argmax}} F + \pi_1 U_{ave} - \pi_2 U_{var} - \pi_3 \gamma \\
s.t. &\ \ G', R, C_{ij}\ (i,j)\in E' \\
&\ \ l_{max}, k \in \mathbb{Z}^+
\end{split}
\end{equation}

Note that the above optimization problem is distinct from the multi-commodity flow formulation for optimal routing on traditional networks. Requested demands $\bar{f}^r$ do not serve as the constraints. The objective function is also flexible and contains multiple terms, instead of using the throughput or delay as the sole objective value. Depending on the computational capacity of the scheduler (the central processor), this optimization task could be solved either by brute force (given the relatively constrained degree of freedom), i.e. searching over $\{l_{max}, k, \alpha, \beta\}$ through simulations, or by applying some machine learning techniques.
One could, e.g., use supervised learning or reinforcement learning with the weighted adjacency matrix for the realized graph $G'$ and requests $R$ as input , and for output the learned optimal parameters $X^*$, with routing performances evaluated and recorded, either in a hard manner (i.e., success/fail; for supervised learning), or in a soft manner (i.e., as rewards; for reinforcement learning). 
The advantage of using machine learning approaches is that one could deal with multiple terms in the objective function more flexibly by designing sophisticated learning workflow. In practice, this would  essentially allow learning high-dimension heuristics to evaluate the routing performance,  beyond the straightforward measures discussed above. Given that a few critical mechanisms in real applications are excluded in the current simulation framework, notably network failures, it is desired that an effective machine learning workflow is assembled for the parameter determination of the routing scheme; we leave this open question for future work.

\section*{Computational Complexity}

Before presenting  the simulation results, here we  briefly discuss  the computational complexity of the main workflow, namely,  \textit{Steps 2-4}. 

\textit{Step 2} incurs the largest computation cost in determining the shortest paths, with a computational time  $O(|R|kV^3) = O(|R|kE^{1.5})$ for square lattices.  \textit{Step 4} demands a cost $O(|R|k)$ to calculate the allocated capacity for each request path.  
The complexity of the allocation process in \textit{Step 3} depends on the algorithm.  The time is $Q_{PS} = O(E|R| l_{max}) \sim O(E|R|k)$ for \textit{proportional share} since $l_{max}\sim k$, while it is $Q_{PF} = O(C_0)$ for \textit{progressive filling}. 
The computational complexity of \textit{propagatory update} is higher due to its iterative nature. In the worst case, the max flow on each path will decrease from the maximum value $C_0$ to the minimum value $f_{min}$ step by step. Therefore, for $|R|$
 requests 
and $k$ paths for each request, the no improvement counter will be reset to 0    $|R|k(C_0 - f_{min})$ times, resulting in a computational cost $O(|R|k(C_0 - f_{min})E)$. Since $C_0 - f_{min} \sim C_0$, the time complexity of the \textit{propagatory update} algorithm is approximately:
\begin{equation}
Q_{PU} \sim O(E|R|k C_0) = C_0 Q_{PS},
\end{equation}
 The \textit{propagatory update} method is then the slowest among the three scheduling schemes in general. Overall, the time complexity $Q_{alg}$ of the routing scheduler is:
\begin{equation*}
Q_{alg} = O(E^{1.5}|R|k) + O(E|R|k) + O(|R|k) = O(E^{1.5}|R|k)
\end{equation*}
for \textit{proportional share}, $O(E^{1.5}|R|k+C_0)$ for \textit{progressive filling}, and $O((E^{0.5} + C_0)E|R|k)$ for \textit{propagatory update}.

\section*{Simulation Results}
We performed extensive simulations  to study the effect of design parameters $X = \{l_{max}, k, \alpha, \beta\}$, and to demonstrate the robustness of the routing design under different assumptions on entanglement fidelity, $F_{mean}$, $F_{std}$, $F_{th}$, and operation success probabilities  $p_{in}$, $p_{out}$. We characterize the performance of the simulation outcomes   from multiple perspectives, using  the established metrics: the throughput (total flow $F$), the traffic   ($U_{ave}$ and $U_{var}$), the delay (path stretching factor $\gamma$), and fairness over requests (Jain's index $J_{req}$) and paths ($J_{path}$). 
We focus on the comparison among the three scheduling schemes (PS, PF, PU in short). 
To limit the simulation complexity and better understand system parameters, we consider an $8\times8$ square lattice, with two connection requests $[s_{i}, t_{i}], i=1,2$ with  same weight and identical request distance. Here the request distance is defined as the  Manhattan distance along one direction (the Manhattan distance along the two directions is same in our simulations for simplicity).   
Under this baseline case, we first demonstrate a sample result of our general routing scheme. Then we study the choice of $k$, which is used in all three scheduling schemes; tests show that it is the most important design parameter in determining the performances of the three algorithms. 
Later we study the other three parameters $\{l_{max}, \alpha, \beta\}$ used in PS and PU, with a focus on the fairness of routing  as compared to PF which guarantees the max-min fairness. Finally we study the system's robustness under different conditions of quantum parameters. For each set of parameters, 200 simulation runs are performed and the results are averaged to account for the stochastic nature in the quantum setup. 
Beyond the baseline case scenario, alternative network topologies (hexagonal/triangular), arbitrary connection requests (varied request number $|R|$ and arbitrary distance) and potential network failures (edge/node failure) are studied in further simulations.

\begin{figure*}[ht]
\centering
\includegraphics[scale=0.4]{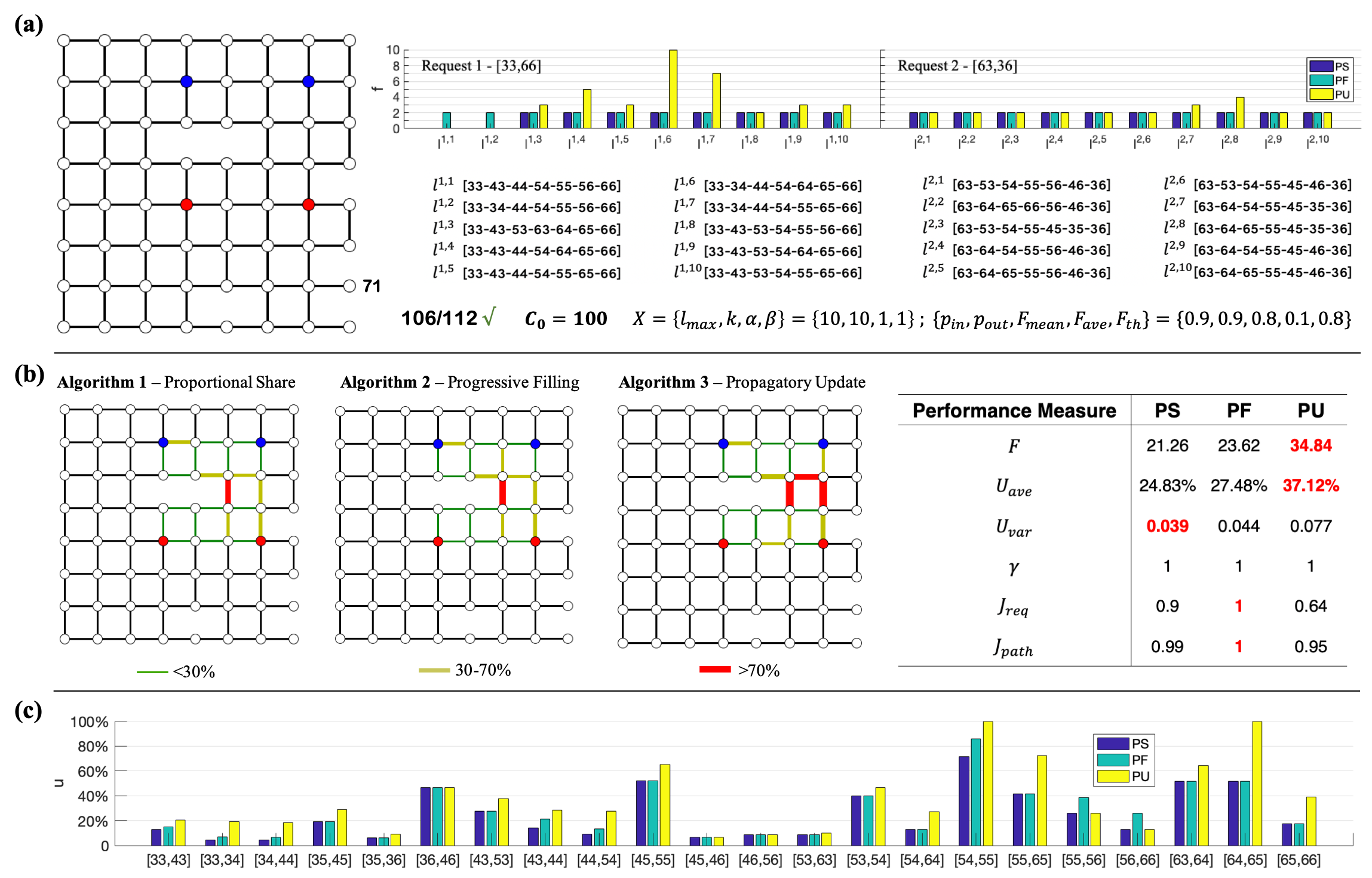}
\caption{\label{BaseCase} Baseline case results.  \textbf{a.} (left) Network topology after entanglement purification; (right) 10 connection paths for each request (bottom) and the allocated flow for each path under the three scheduling schemes (top). Here the nodes are represented by a two digit number and the bottom left corner is denoted as 00. An example of node 71 is labeled in the graph plot.  \textbf{b.} Comparison of routing performance. Traffic plots (left) and performance measures (right). We show the traffic plot in a 8-by-8 network where edge color and width represent the capacity utilization of the corresponding edge ($<30\%$ in green, $30\%-70\%$ in yellow and $>70\%$ in red, large width corresponds to larger utilization).  
\textbf{c.} Capacity utilization on (utilized) edges under the three schemes.  For simplicity sake, in all simulations hereafter we set the initial capacity of each edge to be $C_{0} =100$, and the fidelity distribution among edges follows a normal distribution with mean  $F_{mean} = 0.8$ and standard derivation $F_{std} = 0.1$.}
\end{figure*}

\subsection*{Baseline case} In Fig.~\ref{BaseCase}, we demonstrate a sample result of our routing scheme. Initially, $C_0 = 100$ entangled pairs are prepared at each station. After entanglement purification (\textit{Step 1}), 106/112 edges in the $8\times8$ square lattice are maintained, under quantum parameters $\{F_{mean}, F_{std}, F_{th}, p_{in}, p_{out}\}$ = $\{0.9, 0.9, 0.8, 0.1, 0.8\}$ (left, Fig.~\ref{BaseCase}~(a)). Two connection requests ($[33,66],[63,36]$) are received, with source/terminal nodes shown in red/blue; the first/second coordinate indicates the horizon/vertical axis (see node 71 in Fig.~\ref{BaseCase}~(a)). System parameters are $X = \{l_{max}, k, \alpha, \beta\} =  \{10, 10, 1, 1\}$. Ten paths are identified for each request (\textit{Step 2}); after  \textit{Step 3} and \textit{Step 4}, the routing results under three scheduling schemes are compared. Fig.~\ref{BaseCase}~(b) shows the traffic on the network in a graphic view (left). We compare performance measures of the three scheduling algorithms: PU obtains the largest throughput, and it best exploits edge capacities; indeed, it shows that the scheduled flows on paths are almost always the highest on PU (right top, Fig.~\ref{BaseCase}~(a), so are the capacity utilizations on different edges (Fig.~\ref{BaseCase}~(c) ). PS obtains the smallest variance in edge capacity utilizations, suggesting that the routed traffic is more evenly distributed in PS than in PU or PF; yet PS has the poorest throughput $F$ and average capacity utilization $U_{ave}$. PF has less throughput and capacity utilization than PU, but as expected, the scheduling is completely fair, with respect to both requests ($J_{req}$) and paths ($J_{path}$). PU is the least fair scheme among the three and it can be attributed to the large flow on few paths (e.g., $f_{1,6}, f_{1,7}, f_{2,8}$ in Fig.~\ref{BaseCase}~(a)).
Various tests are simulated under different parameter sets, and results suggest that the baseline case is representative of the three schemes' performance; the comparison of performance in Fig.~\ref{BaseCase}~(b) largely hold in an aggregated sense.

\subsection*{Dependence on request distance and choice of $k$}

\begin{figure*}[ht]
\centering
\includegraphics[scale=0.4]{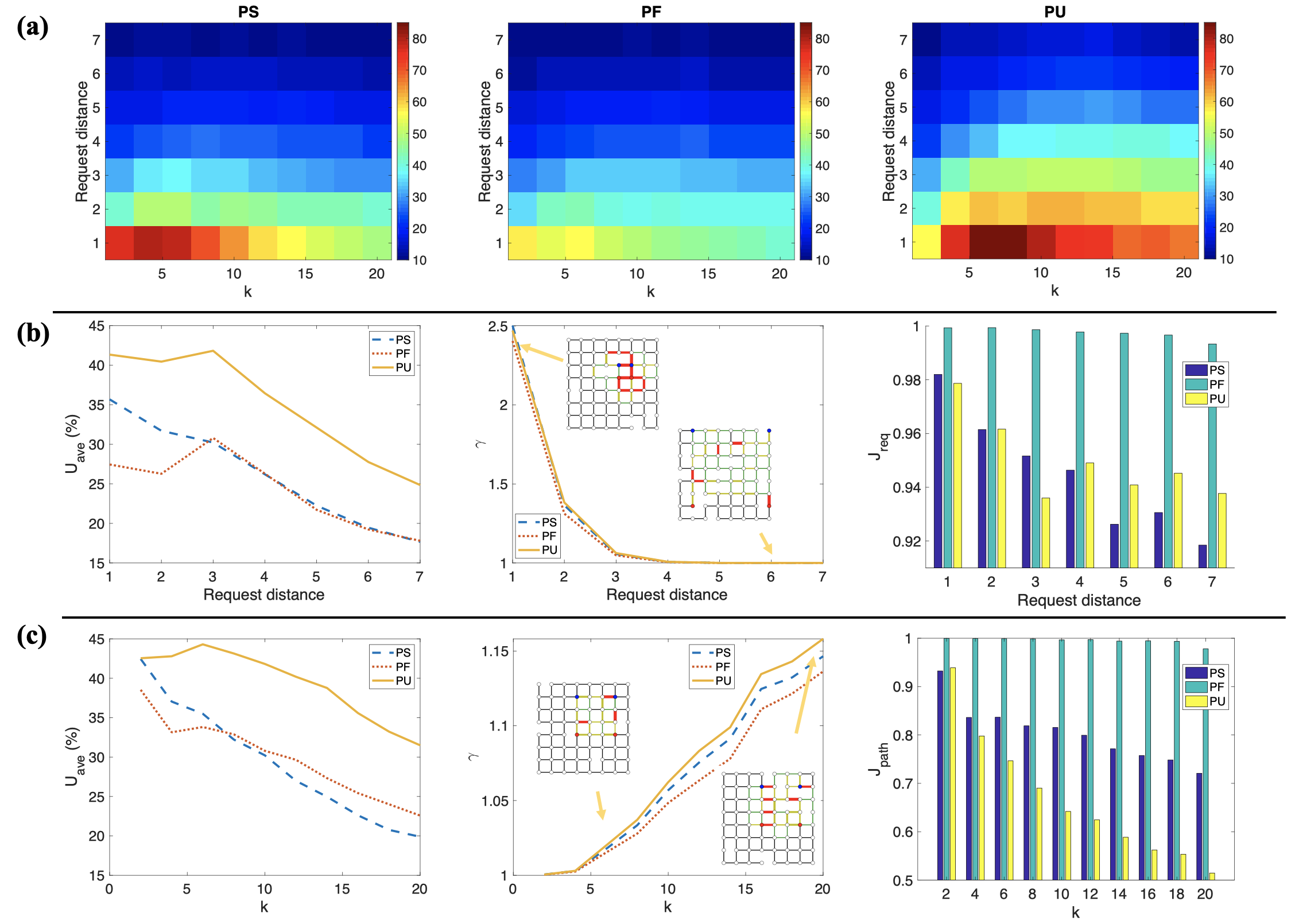}
\caption{\label{Fig4_dis_kmax} Algorithm performance as a function of number of shortest paths $k$ and request distance. \textbf{a.}  Heat map of the throughput $F$ for the two requests. \textbf{b.} Routing performance with respect to request distance at $k=10$. $U_{ave}$, $\gamma$ and $J_{req}$ plots are shown here. \textbf{c.} Routing performance with respect to k at request distance 3. $U_{ave}$, $\gamma$ and $J_{path}$ plots are shown here.  Examples of traffic plots with different parameters for the PU algorithm are shown in the inset of $\gamma$ plots. The other system parameters are $\{l_{max}, \alpha, \beta\} =  \{15, 1, 0\}$. }
\end{figure*}

In Fig.~\ref{Fig4_dis_kmax}~(a) we show the total flow $F$ of the three algorithms. The throughput, as well as the minimum flow of the two requests, decay exponentially as the distance between request pairs grows. The PU algorithm results in obviously larger throughputs (larger number of entangled pairs), while the other two algorithms are comparable to each other. We notice that the throughput will reach a maximum at moderate $k$ values. This can be explained as slightly higher $k$ values provide more freedom in generating large flows, but beyond the optimum  $k$ point  there will be more paths along each edge on average, leading to congestion on the utilized edges and resulting in more bottlenecks for each path thus decreasing the throughput. 
We also observe that with large request distance, $k$ has less of an effect, since there are enough routes to select and less edges where congestion can occur.

Fig.~\ref{Fig4_dis_kmax}~(b) shows the behavior of the other metrics with respect to request distance. As the number of paths between request pairs becomes exponentially large with the request distance, for fixed $k$ the traffic is spread out on all edges of these paths, leading to decreasing capacity utilizations. Similarly, at large request distance, there is no need for the flows to take detours and all utilized paths are indeed shortest paths which yields  $\gamma = 1$. This is clearly depicted in the traffic plots. We also observe that the PU method is superior in utilizing the edge channels and has better capacity utilizations due to information propagation among different edges during its iterative process. All the algorithms demonstrates excellent performance in dealing with multiple requests. In particular, the PF algorithm, due to its step-by-step filling nature, performs excellently in balancing the requests and paths.

For fixed requests as shown in Fig.~\ref{Fig4_dis_kmax}~(c) , choosing a larger $k$  means that the routing algorithms tend to explore more edges,  which results in a lower average capacity utilization. On the same note, circuitous paths will be utilized in routing, which increases the normalized path length thus the time delay.  Meanwhile, for the PS and PU algorithm, the flow variance among different paths will increase, leading to decreasing fairness over paths. These results point out that the parameter $k$ is crucial in determining path properties.

\begin{figure}[ht]
\centering
\includegraphics[scale=0.22]{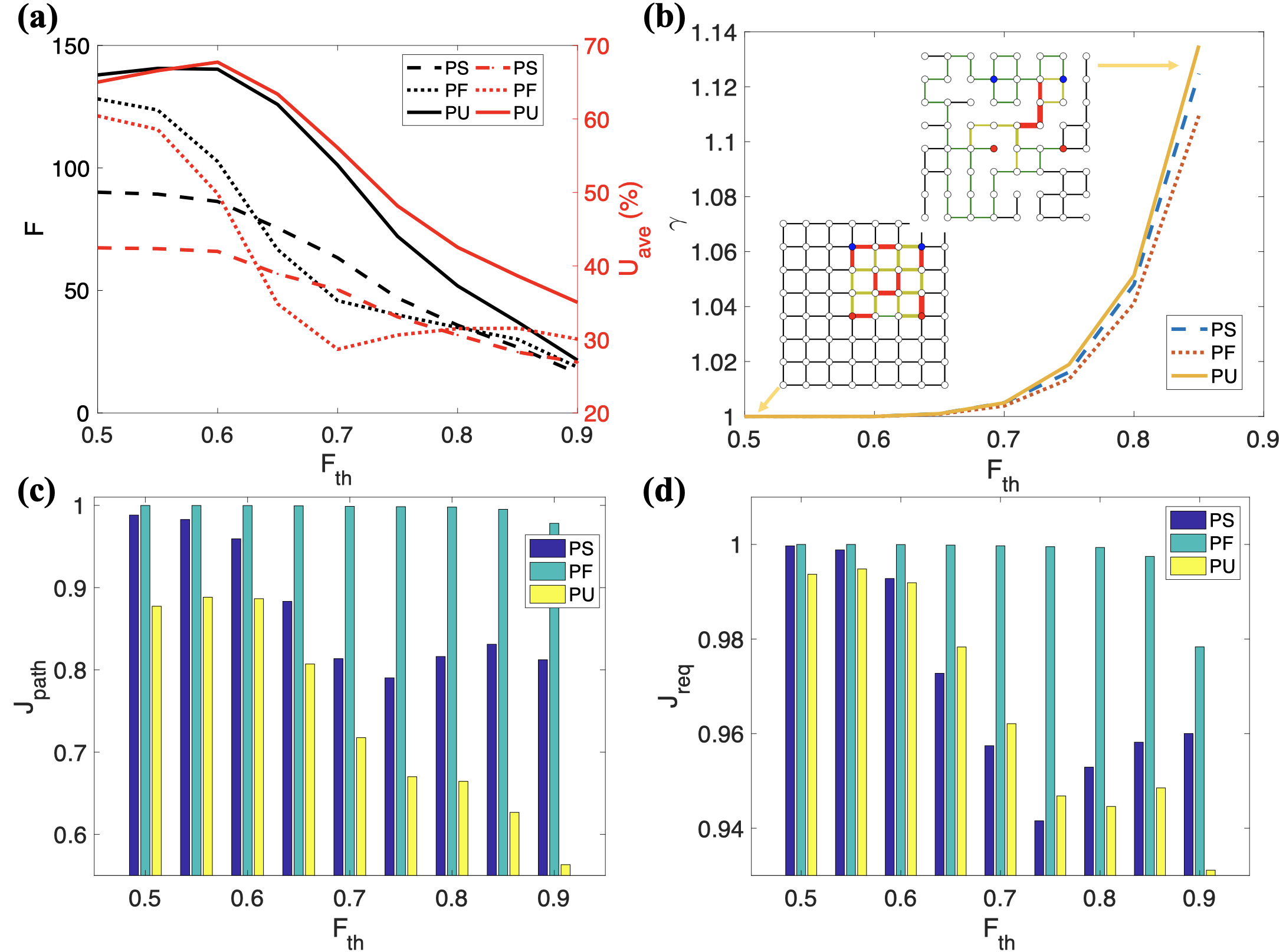}
\caption{\label{Fig6_fidelity} Protocol performance with respect to fidelity threshold. \textbf{a.} Total flow and average capacity utilization. Inset shows the minimum flow of the two requests. \textbf{b.} Normalized path stretching factor. Traffic plots at two extreme case for the \textit{propagatory update} algorithm are presented. \textbf{c-d.} Fairness over paths (c) and requests (d). The fidelity distribution among edges follow the normal distribution $N(0.8,0.1)$ and the request distance is fixed to be 3 here. The other system parameters are $\{l_{max}, \alpha, \beta\} =  \{15, 1, 0\}$. }
\end{figure}

\subsection*{Effect of fidelity threshold}
Finally, we discuss the fidelity threshold that can affect the entanglement purification steps (thus the realized capacity $C_{ij}$). 
 Specific tasks in quantum network would demand different error rates or fidelity thresholds and the corresponding entanglement purification will modify the topology of the network in different ways. 
In Fig.~\ref{Fig6_fidelity} we show the dependence of the protocol performance on fidelity threshold. As the fidelity bound raises, more purification steps are needed, monotonously decreasing the  edge capacity and hence the throughput. Intuitively, one might expect that for high fidelity threshold,
in general edges will have less realized capacity $C_{ij}$ which can be fully utilized, leading to large capacity utilization.
However, as shown in Fig.~\ref{Fig6_fidelity}~(a), the average capacity utilization follows the same trend as the throughput. This is due to the fact that for fixed $k$, the edge removal due to repeated purification steps will force the routing scheme to explore more edges in the network  thus decreasing the overall capacity utilization. This is clearly shown in the traffic plots (Fig.~\ref{Fig6_fidelity}~(b)). Similarly,  for high fidelity threshold, the normalized path length is clealy larger than unity for the three algorithms and the flow among different paths tends to have large variance, thus small fairness over paths for PS and PU (Fig.~\ref{Fig6_fidelity}~(d)). Nevertheless, the two requests are well balanced with high fairness over request for all allocation algorithms (Fig.~\ref{Fig6_fidelity}~(d)).

\subsection*{Potential network failures} 

\begin{figure}[ht]
\centering
\includegraphics[scale=0.18]{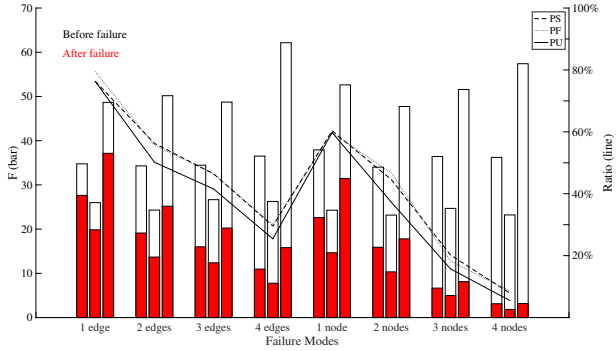}
\caption{ \label{Robustness} Robustness of routing results against network failures. Comparing system throughput before and after the failure of the three scheduling schemes. Different failure modes (1-4 edges failure, 1-4 nodes failure) are simulated.}
\end{figure}

One critical feature of our routing scheme is to be multi-path, and the claim is that by utilizing more than one path, the routing will be more robust to potential network failures. We simulate both edge failures (an edge is dead) and node failures (a station is dead; all edges linked to it are dead) to test the scheme's robustness (Fig.~\ref{Robustness}). The system throughputs $F$ before and after the failure are compared for each scheduling scheme, under different failure modes (1-4 edges failure, 1-4 nodes failure; edges and nodes are chosen from those being utilized). 20 simulations are averaged at each data point under baseline case parameters (see Figure \ref{Robustness}). Results show that, as expected, as the failure mode gets more and more serious, the system  achieves smaller and smaller throughput, but is still working; after the shutdown of 4 out of the 16 utilized stations, the system still maintains some outputs. We  also notice that of the three schemes, PS shows the greatest robustness, slightly greater than PF, whereas PU's superiority in throughput $F$ gradually vanishes in face of network failures. Similar robustness conditions are identified on other performance measures (see Supplemental Material).

\subsection*{Number of requests per window}
\begin{figure}[ht]
\centering
\includegraphics[scale=0.19]{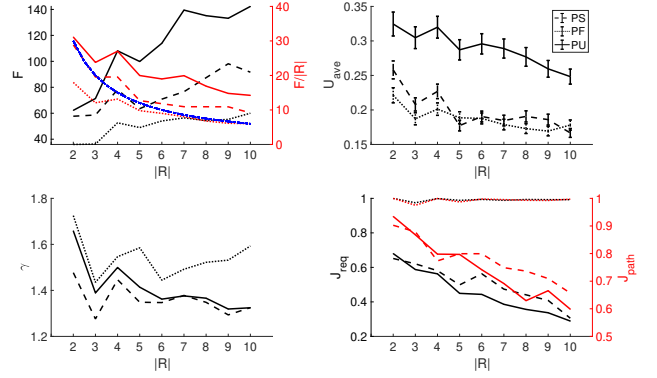}
\caption{ \label{RequestNum} Performance under different number of requests (2-10) per window. (top left) Throughput $F$ and normalized throughput $F/|R|$; blue curve shows the reference line $60/|R|$. (top right) Traffic mean $U_{ave}$ and variance $U_{var}$ (error bar; reduced scale of factor 2). (bottom left) Delay $\gamma$. (bottom right) Fairness $J_{req}$ and $J_{path}$.}
\end{figure}

Within each processing window, we simulate 2-10 random connection requests (arbitrary $[s,t]$ pairs)  and compare routing performances (Fig.~\ref{RequestNum}). 20 simulations are averaged at each data point under based case parameters (see Figure 3). With $|R|$ increasing, the system becomes more crowded: $F$ goes up whereas $F/|R|$ goes down, but the degradation curvature is better than inverse-proportional (blue line); the superiority of PU in throughput is always maintained. Meanwhile, the network traffic is more diluted, and the routing less fair for PU and PS, with $U_{ave}$, $U_{var}$, $J_{req}$ and $J_{path}$ falling quasi-linearly. There is no consistent change in average routing delay $\gamma$; PU seems to produce less delayed routing under more requests, but the result is not significant.
 
\section*{Discussion}

We  benchmarked the performance of the routing scheme under different parameter regimes. Here we highlight differences in resource allocation for the three scheduling schemes. 
Performance results show that the three schemes have different advantages: \textit{progressive filling} is the fairest method, whereas the fairness of the two proportional methods is compromised to a non-trivial extent, and  depends on the system parameters $\alpha, \beta$. 
In terms of system throughput and capacity utilization, the new \textit{propagatory update} scheme stands out by a large margin, and the superiority is largely maintained under all network conditions, even in face of network failures. However, as a tradeoff, the increased throughput are not attributed to all requests in a fair manner, and \textit{propagatory update} routing results demonstrate the largest variance on edge utilizations as well. The least efficient scheme is \textit{proportional share}, which generates the least throughput. However, the variance of edge utilization is smaller than in \textit{propagatory update}, and it is more robust against network failures. This suggests that this scheme utilizes network resources in the most economical way. Also, since a global information table is not maintained (hence not subject to errors) and proportional allocations are performed only at the local scale, the \textit{proportional share} may also be the most robust scheme against network errors on top of failures.

We now proceed to present the scope of use, limitations and potential future extensions of the routing scheme.
We have demonstrated the performance of our routing scheme with a $8\times8$ square lattice network with edge capacity $C_0 = 10^2$ and $<10$ requests per time window. In reality, the implementation scale of quantum networks will be determined by various physical, geological and economic constraints. Current lab-scale quantum networks are very small ($\sim 2\times2$) with very limited edge capacity $C_0$ (only few qubits), since the preparation of entangled pairs is extremely costly; for our routing scheme to be put in use, huge leaps in quantum engineering are necessary. Practically, the lattice size and edge capacity should be decided by the specific functionality of the quantum network, and the number of requests being processed within a time window are determined by service considerations. Also note that alternative lattice topologies (e.g., hexagonal, triangular) could be experimented in real applications; current results apply to square lattice, yet it is believed that similar results could be obtained on other common regular lattice structures.

We further note that, in the space domain, we are assuming a central processor in the network and global link-state information is broadcast to each nodes. 
Under restricted cases, however, only local link-state knowledge (link with the nearest neighboring nodes) is available at each node. For example, as the size of the network $G(V,E)$ increases, the classical communication time is non-negligible and can be far beyond the quantum memory lifetime.  In this scenario, instead of finding the  paths with a global processor, one can perform entanglement swapping on two local quantum memories in a way such that the sum of their distance to the request pair can be minimized, as shown in \cite{MPant2019npj}. However, in a more complex quantum network where multi-request, multi-path and multi-channel are involved, searching for an efficient routing strategy remains elusive and might be interesting for future study.

In the time domain, the current protocol is used for processing a certain number of connection requests within one active time window; it is expected that a queueing model \cite{gross2008fundamentals} might be constructed to tackle the network functioning  beyond the processing time window. 
Such queueing system should provide guidelines for coordinating the reception of incoming connection requests with the recycling of entangled pairs, in which case, processing windows could essentially overlap and one does not need to wait for the end of one batch of requests  before starting the next window. In this scenario, other concerns in routing design might  arise, for example,  dealing with network congestions, e.g. subsequent re-distribution of capacities after initial allocation \cite{PhysRevE.73.046108}. Since these issues have been intensively studied in classical network, discussions on such high-level infrastructures of the current protocol are not included in this paper and are left for future studies.

Next we point out that in the above protocol we consider entanglement purification only at the first step, that is, only between adjacent nodes. As successive entanglement swapping operations are imperfect, they might degrade the fidelity of generated remote entanglement. To overcome this drawback, one can adopt a nested protocol \cite{PhysRevA.72.052330,Perseguers_2013} where repeated purification is performed after each entanglement swapping step to ensure that the remote entangled states have regained a sufficiently large fidelity. Then an updated version of the routing scheme in which the resource consumption during successive entanglement purification is taken into account might be designed.

Finally, while here we consider entanglement between two nodes, i.e., Bell states, an extension to GHZ states and local GHZ projection measurement might bring advantages in generating remote entanglement \cite{Pirker_2018,Dahlberg_2018,Pirker_2019,Hahn_npj2019}. 
For example, even if the GHZ states are more difficult to prepare and more fragile to environment noise, they provide  more configurations  on a quantum network, thus creating more complex topology which can be unreachable for Bell states. The compromise between this gain and its cost can in principle benefit our entanglement routing designs. Similarly, the GHZ projection measurement in a local station can glue more than two remote stations together and modify the network topology. The routing strategy that utilize these operations will be an interesting topic to study, both mathematically and physically.

\section*{Conclusion}
 In conclusion, we have proposed an efficient protocol that tackles the routing problem on quantum repeater networks. We have shown that entanglement purification, which increases entanglement fidelity, can  modify the network topology and affect the following routing design. For general \textit{multi-request}, \textit{multi-path} and \textit{multi-channel} problems, we developed three algorithms that can allocate limited resource to efficiently generate entangled pairs between remote station pairs. 
Among the first attempts to tackle this complex problem, our routing scheme  provides a solution to exploit the full capability of complex quantum networks, where efficient entanglement generation of remote station pairs are desired. The routing design here may also inspire future work that combine the quantum information science with classical network theory.

\acknow{We thank Eytan Modiano for helpful discussions.   }

\showacknow{} 

\bibliography{Qnetwork_arXiv}

\newpage


\section*{Supplementary material}

# Supplemental material for "Effective routing design for remote entanglement generation on quantum networks"

Changhao Li[a,1], Tianyi Li[*c,1], Yi-Xiang Liu[a,b], and Paola Cappellaro [†a,b,2]

[a] *Research Laboratory of Electronics, Massachusetts Institute of Technology, Cambridge, MA 02139*
[b] *Department of Nuclear Science and Engineering, Massachusetts Institute of Technology, Cambridge, MA 02130*
[c] *System Dynamics Group, Sloan School of Management, Massachusetts Institute of Technology, Cambridge, MA 02139*
[1] *The authors contribute equally to the work.*

# Contents

# 1 Entanglement swapping and purification

We first briefly review the basic elements of entanglement swapping and purification for the convenience of readers.

## 1.1 Entanglement swapping

Due to the optical loss during information transmission, entanglement swapping would be essential in generating long distance entanglement. Here we consider memory-assisted quantum repeaters, as shown in Fig. S1. The quantum memory 2 and 3 are in the same station hence local bipartite operations can be performed at low cost. At the beginning, we create Bell states between memory 1 (3) and 2 (4), the state of which can be written as:

$$|\psi^-_{12}\rangle|\psi^-_{34}\rangle = \frac{1}{2}(|\psi^+_{14}\rangle|\psi^+_{23}\rangle - |\psi^-_{14}\rangle|\psi^-_{23}\rangle - |\phi^+_{14}\rangle|\phi^+_{23}\rangle + |\phi^-_{14}\rangle|\phi^-_{23}\rangle) \tag{S1}$$

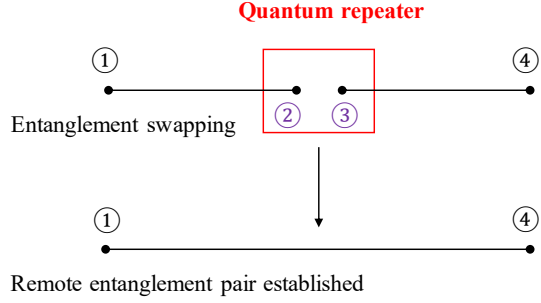

Figure S1: Illustration of quantum repeaters. The black dots (lines) represent quantum memories (optical links).

Then we can see that a joint Bell state measurement on memory 2 and 3 will project memory 1 and 4 into Bell states. For example, the measurement result of $|\psi_{23}^-\rangle$ would reveal that we create a non-local entangled states $|\psi_{14}^-\rangle$. We point out that with quantum repeaters, the time needed to generate remote entanglement (with certain fidelity) would scale polynomially as a function of distance.

## 1.2 Entanglement purification

Now we introduce the principle of entanglement purification with a simple example. We consider two pairs of qubits shared by Alice and Bob, the state of which is described by the density matrix $\rho_{AB}^j = f|\Psi^+\rangle\langle\Psi^+| + (1-f)|0\rangle\langle 0|$, where $j$=1,2 and $|\Psi^+\rangle$ is the state to distill. By applying the local C-NOT gates between the two qubits located at Alice (Bob), we can find the new density matrix:

$$\rho = f^2|\Psi^+\rangle_1\langle\Psi^+| \otimes |\Phi^+\rangle_2\langle\Phi^+| + \tag{S2}$$
$$(1-f)^2|00\rangle_1\langle 00| \otimes |00\rangle_2\langle 00| + \tag{S3}$$
$$f(1-f)|00\rangle_1\langle 00| \otimes |\Psi^+\rangle_2\langle\Psi^+| + \tag{S4}$$
$$\frac{f(1-f)}{2}(|0101\rangle + |1010\rangle)_{1,2}(\langle 0101| + \langle 1010|) \tag{S5}$$

By performing measurements on the two target qubits, the state of pair one will become $\rho_{AB}^1 = \frac{f^2|\Phi^+\rangle\langle\Phi^+|+(1-f)^2|00\rangle\langle 00|}{f^2+(1-f)^2}$ conditioned on the same measurement result (00 or 11) of pair 2 (with success probability $P = f^2 + (1-f)^2$ ). The fidelity of state $|\Psi^+\rangle$ after the $i$-th distillation will be $f^i = \frac{(f^{i-1})^2}{(f^{i-1})^2+(1-f^{i-1})^2}$. As shown in Fig. S2, if the initial $f^{i-1}$ is above 0.5, highly entangled states can be generated. Note this protocol only involves inexpensive local operations. However, the purification of state fidelity is at the cost of entanglement generation rate (since $P < 1$) and multiple entanglement pairs.

In the limit of $1 - f_i \ll 1$, the purification curve will simply be $1 - f_{i+1} = (1 - f_i)^2$ while the success probability is $f_i$. To obtain one EPR pair with the desired threshold $F_{th}$, the average number of fresh EPR pairs required will be:

$$n_r = 2^m \frac{1}{\prod_i^m P_i}, m = \frac{log(1-F_{th})}{2log(1-\langle F \rangle)} \tag{S6}$$

where $\langle F \rangle$ is the initial EPR fidelity and m is the number of purification steps one needs to perform.

---

*tianyil@mit.edu
†pcappell@mit.edu

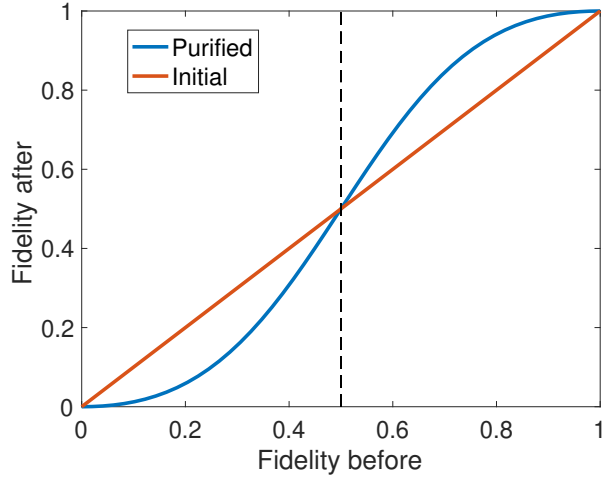

Figure S2: Purification curve. Above the 0.5 threshold, the purification process will gain fidelity of desired state.

## 2 Capacity allocation scheduling

In *Step 3* of the routing protocol, capacity allocation scheduling is carried out. We propose and compare three methods for this task: (i) *proportional share*, (ii) *progressive filling* and (iii) *propagatory update*, of which (i) and (ii) are based on well-known scheduling heuristics on classical networks and (iii) is our invention.

### 2.1 Algorithm 1: *Proportional Share*

The difference of the *proportional share* and the *propagatory update* methods lies in the direction of managing capacity allocation. The first method forwardly divides the realized (maximum) edge capacity $C_{i,j}$ into proportions, according to the weights of divisors. At the first stage, the division is carried out on different requests. On edge $(i,j)$, the capacity allocated for request $r$ is given by:

$$C_{i,j}^r = C_{i,j} \frac{w_r (\#r_{i,j})^{-\beta}}{\sum_r w_r (\#r_{i,j})^{-\beta}}, \tag{S7}$$

where $\#r_{i,j}$ is the number of paths under request $r$ passing edge $(i,j)$, and $w_r$ is the weight of request $r$. We assume that the allocated capacity is proportional to the request weight, but is inversely proportional to the power ($\beta$) of the number of paths under the requests, i.e. $\beta > 0$; requests have fewer paths across the edge are compensated.

At the second stage, in a similar top-down manner, the allocated capacity on each request is further divided among different paths. The flow capacity for path $l$ of request $r$ on edge $(i,j)$ is given by:

$$C_{i,j}^{r,l} = C_{i,j} \frac{d_{r,l}^{\alpha}}{\sum_l d_{r,l}^{\alpha}}. \tag{S8}$$

Similarly, we assumes that the division is proportional to the power ($\alpha$) of the path length $d_{r,l}$. For $\alpha = 1/-1$, allocation is proportional/inverse proportional to the path length; for $\alpha = 0$, no dependence

3

occurs. The polarity of $\alpha$ represents different strategies for the coordination of allocation on different paths: $\alpha > 0$ corresponds to the situation where longer paths get more flow quota as compensation, since long paths have low success rates; on the contrary, $\alpha < 0$ means shorter paths get more flow quota, as a strategy of prioritizing short paths as major paths for requests. $\alpha$ and $\beta$ could be viewed as controlling the exploitation-exploration tradeoff of the algorithm, similar to the idea in the ant colonization algorithm. When $\alpha > 0$, the algorithm encourages the exploration of different paths by compensating longer paths, while $\alpha < 0$ indicates the exploitation on the usage of short paths. Similarly, when $\beta > 0$ the edge capacity is distributed on multiple requests in a fair manner, and $\beta < 0$ is the situation that prioritizes those requests with more paths passing through the edge. The optimal $\alpha$ and $\beta$ is experimented in simulations and the corresponding strategies are discussed.

Finally, the allocated capacity for each path is lower-capped by the minimum flow $f_{min}$:

$$C_{i,j}^{r,l} = max\ (C_{i,j}^{r,l}, f_{min}) \tag{S9}$$

as a hard constraint. As expected, this treatment will likely raise the problem of capacity outbound, where:

$$\sum_{r,l} C_{i,j}^{r,l} > C_{i,j}. \tag{S10}$$

If this happens, the overflow $\sum_{r,l} C_{i,j}^{r,l} - C_{i,j}$ is iteratively subtracted from paths receiving the largest share, until the outbound is resolved, which is guaranteed to converge under the definition of $f_{min}$. This forward local allocation process is summarized in Algorithm 1.

---

**Algorithm 1** (*Proportional Share*) Capacity Allocation Scheduling

    **for** $(i,j)$ in $E$ **do**
        Cut path list $L = \{L_{i,j}^r\}$ s.t. $|L| \leq l_{max}$ by path length. Keep short paths/sole paths for requests.
        **Stage 1** Allocate $C_{i,j}$ onto different requests $R_{i,j} = \{r \text{ in } h \text{ for } h \in H\}$.
        **for** $r$ in $R_{i,j}$ **do**
            $C_{i,j}^r = C_{i,j} \frac{w_r/(\#r_{i,j})^\beta}{\sum_r w_r/(\#r_{i,j})^\beta}$
        **end for**
        **Stage 2** Allocate $C_{i,j}^r$ onto different paths $L_{i,j}^r = \{l \text{ in } h \text{ for } h \in H\}$ of request $r$.
            **for** $l$ in $L_{i,j}^r$ **do**
                $C_{i,j}^{r,l} = C_{i,j} \frac{d_{r,l}^\alpha}{\sum_l d_{r,l}^\alpha}$
                $C_{i,j}^{r,l} = max\ (C_{i,j}^{r,l}, f_{min})$
            **end for**
            **while** $\sum_{r,l} C_{i,j}^{r,l} > C_{i,j}$ **do** Subtract $C_{i,j}^{r,l} - C_{i,j}$ from $max\ (C_{i,j}^{r,l})$
            **end while**
    **end for**

---

## 2.2 Algorithm 2: *Progressive Filling*

The *progressive filling* method could guarantee max-min fairness. Unlike the other two methods, in the *progressive filling* algorithm, all paths are treated equally. For each path $l$ of each request $r$, flow

4

$f^{r,l}$ starts at zero. The unsaturated edge set $E_{un}$ is the entire $E$, and the unsaturated path set $L_{un}$ is all paths $L^r$ for all $r \in R$.

The filling process assigns a uniform increase on each unsaturated path. Since in our setting edge capacities are integers, the increment is simply 1. Edges are gradually saturated along the filling process, and paths utilizing saturated edges stop flow increases. The integer scenario decides a slight different definition of edge saturation from the definition for non-integer flows, since it is often the case that an edge is left with a small number of unallocated channels, yet insufficient to distribute one more channel on all the unsaturated paths through it, i.e, $C_{i,j} - \sum_{r,l} f_{temp}^{i,j;r,l} < |H|_{i,j} - |L|_{un}^{i,j}$; in this case the edge is viewed as saturated (with a small leftover capacity).

Therefore, at each time step, an attempt of increment 1 is made on all $l \in L_{un}$. On each edge, we check the capacity constraint and decide if the attempt is successful, i.e., if the attempted aggregated flow is out of bound. When the attempt is successful, if the remaining capacity after deducting the attempted aggregated flow is less than the number of unsaturated paths passing through it, the edge is saturated and removed from $E_{un}$. Correspondingly, paths on this edge are removed from $L_{un}$. The filling process terminates when the new attempt is unsuccessful, i.e. it is not possible to assign one more channel uniformly to all unsaturated paths, without causing at least one remaining unsaturated edge to go off bound. Note that since flows increase in a fair manner, $f_{min}$ is unnecessary for this method.

---

**Algorithm 2** (*Progressive Filling*) Capacity Allocation Scheduling

1: Initialize flow matrix $f^{r,l}$ for each path $l$ of each request $r$.
2: Initialize unsaturated edge set $E_{un} = E$; unsaturated path set $L_{un} = \{L^r\}$ for $r \in R$.
3: **while** True **do**
4:     $end\ flag \leftarrow 0$
5:     **Stage 1** Attempt an increment addition of flow uniformly on all unsaturated paths.
6:         **for** $l \in L_{un}$ **do** $f_{temp}^{r(l),l} = f^{r,l} + 1$
7:         **end for**
8:     **Stage 2** Check the capacity constraint on each edge.
9:         **for** $(i,j) \in E$ **do**
10:            **if** $\sum_{r,l} f_{temp}^{i,j;r,l} > C_{i,j}$ **then**
11:               $end\ flag \leftarrow 1$
12:               **break**
13:            **end if**
14:            **if** $C_{i,j} - \sum_{r,l} f_{temp}^{i,j;r,l} < |H|_{i,j} - |L|_{un}^{i,j}$ **then**
15:               Remove $(i,j)$ from $E_{un}$.
16:            **end if**
17:         **end for**
18:     **if** $end\ flag = 1$ **then**
19:         **break**
20:     **end if**
21:     **Stage 3** Record succeeded attempt.
22:         $f^{r,l} \leftarrow f_{temp}^{r,l}$ for all $r, l$
23:         Update $L_{un}$ according to updated $E_{un}$.
24: **end while**
25: Obtain flow matrix $f^{r,l}$.

## 2.3 Algorithm 3: *Propagatory Update*

In the *proportional share* method, each edge decides its own capacity allocation and no communication between edges takes place. In the end, the final flow of the path is determined at *Step 4*, where the short-board constraint is applied on each edge. Intrinsically, this method will result in certain waste of edge capacity: capacity on edges will not be fully utilized, if some edges along the same path have narrower channels. Therefore the average capacity utilization of the network will be low.

This problem could be overcome if the short-board constraint is integrated into the capacity allocation process, instead of being carried out at the end. To achieve this, we store the global maximum flow information for each path, and constantly update this information (reducing the maximum flow of each path) throughout the allocation process. In fact, the allocation process on each edge could now be viewed as a backward-moving process: on a specific edge, assume that each requested path tries to claim its maximum flow, and then, we make deductions on those flows based on the current edge capacity. The gap between the sum of (desired) maximum flows over all paths passing the edge and the edge capacity is the amount to be deducted, and now we allocate the capacity deduction (instead of the capacity) to different requests, then to different paths in a similar two-stage manner as in the first method.

The total amount of capacity $\Delta_{i,j}$ to be deducted is given by:

$$\Delta_{i,j} = \sum_{r,l} f_{max}^{r,l} - C_{i,j} \tag{S11}$$

where $f_{max}^{r,l}$ is the desired max flow of path $l$ for request $r$, whose value is set as the flow of the previous edge in the path: $f_{max}^{r,l} = f_{i,j\ prev}^{r,l}$. On each path, the maximum amount of deduction (its maximum possible contribution to the total deduction) $\Delta_{i,j}^{r,l,max}$ is the difference between its maximum desired flow and its minimum flow:

$$\Delta_{i,j}^{r,l,max} = f_{max}^{r,l} - f_{min}. \tag{S12}$$

Correspondingly, the maximum amount of capacity deduction a request could contribute is the sum from all its paths:

$$\Delta_{i,j}^{r,max} = f_{max}^r - f_{min}^r = \sum_l \Delta_{i,j}^{r,l,max}. \tag{S13}$$

At the first stage, the total amount of deduction $\Delta_{i,j}$ is allocated to different requests as:

$$\Delta_{i,j}^r = min(\Delta_{i,j} \frac{\Delta_{i,j}^{r,max}/w_r}{\sum_r \Delta_{i,j}^{r,max}/w_r}, \Delta_{i,j}^{r,max}) \tag{S14}$$

where requests with large weights $w_r$ gets small amount of deduction and the realized amount could not exceed the maximum amount $\Delta_{i,j}^{r,max}$. This allocation procedure begins from requests having the smallest $\Delta_{i,j}^{r,max}$ and proceeds until all $\Delta_{i,j}$ has been allocated, which is guaranteed since:

$$\sum_{r,l} \Delta_{i,j}^{r,l,max} = \sum_{r,l} f_{max}^{r,l} - \sum_{r,l} f_{min} \geq \sum_{r,l} f_{max}^{r,l} - C_{i,j} = \Delta_{i,j} \tag{S15}$$

6

**Algorithm 3** (*Propagatory Update*) Capacity Allocation Scheduling

1: Initialize max flow vector $f_{max}^{r,l}$ for each path $l$ of each request $r$.
2: Set $count = 0$.
3: **while** $count < |E|$ **do**
4:     Select the next edge $(i,j) \in E$ following the fixed edge order with startover.
5:     **if** $R_{i,j} =$ None **then**
6:         $count++$.
7:         **Continue**
8:     **end if**
9:     Cut path list $L = \{L_{i,j}^r\}$ s.t. $|L| \leq l_{max}$ by path length. Keep short paths/sole paths for requests.
10:     Calculate $\Delta_{i,j} = \sum_{r,l} f_{max}^{r,l} - C_{i,j}$, $f_{max}^r = \sum_l f_{max}^{r,l}$, $f_{min}^r = \sum_l f_{min}$
11:     **Stage 1** Allocate $\Delta_{i,j}$ onto different requests $R_{i,j} = \{r$ in $h$ for $h \in H\}$.
12:     Calculate max deduction on requests $\Delta_{i,j}^{r,max} = f_{max}^r - f_{min}^r$
13:     **from** smallest $\Delta_{i,j}^{r,max}$ **for** $r$ in $R_{i,j}$ **do**
14:         $\Delta_{i,j}^r = \Delta_{i,j} \frac{\Delta_{i,j}^{r,max}/w_r}{\sum_r \Delta_{i,j}^{r,max}/w_r}$
15:         **if** $\Delta_{i,j}^r > \Delta_{i,j}^{r,max}$ **then** $\Delta_{i,j}^r = \Delta_{i,j}^{r,max}$
16:         **end if**
17:         Recalculate $\Delta_{i,j} = \Delta_{i,j} - \Delta_{i,j}^r$.
18:     **Stage 2** Allocate $\Delta_{i,j}^r$ onto different paths $L_{i,j}^r = \{l$ in $h$ for $h \in H\}$ of request $r$.
19:     Calculate max deduction on paths $\Delta_{i,j}^{r,l,max} = f_{max}^{r,l} - f_{min}$
20:     **from** smallest $\Delta_{i,j}^{r,l,max}$ **for** $l$ in $L_{i,j}^r$ **do**
21:         $\Delta_{i,j}^{r,l} = \Delta_{i,j}^r \frac{\Delta_{i,j}^{r,l,max}(d_{r,l})^{-\alpha}}{\sum_l \Delta_{i,j}^{r,l,max}(d_{r,l})^{-\alpha}}$
22:         **if** $\Delta_{i,j}^{r,l} > \Delta_{i,j}^{r,l,max}$ **then** $\Delta_{i,j}^{r,l} = \Delta_{i,j}^{r,l,max}$
23:         **end if**
24:         Recalculate $\Delta_{i,j}^r = \Delta_{i,j}^r - \Delta_{i,j}^{r,l}$.
25:     **Stage 3** Apply deduction and determine flow $f_{i,j}^{r,l}$.
26:     $f_{i,j}^{r,l} = f_{max}^{r,l} - \Delta_{i,j}^{r,l}$
27:     **Stage 4** Update max flow $f_{max}^{r,l}$ of the path.
28:
29:     **if** $f_{max}^{r,l}$ updated **then**
30:         Reset $count = 0$.
31:     **else**
32:         $count++$.
33:     **end if**
34: **end while**
35: Determine converged $f_{i,j}^{r,l}$.

i.e. the total amount of desired deduction is no greater than the total amount of maximum deduction.

At the second stage, the amount of deduction on each request is allocated to different paths according

to:
$$\Delta_{i,j}^{r,l} = min(\Delta_{i,j}^r \frac{\Delta_{i,j}^{r,l,max}/(d_{r,l})^\alpha}{\sum_l \Delta_{i,j}^{r,l,max}/(d_{r,l})^\alpha}, \Delta_{i,j}^{r,l,max}). \tag{S16}$$

Again, the maximum constraint $\Delta_{i,j}^{r,l,max}$ is applied. Similar to the first method, the weight of each path is assumed to be proportional to the power $\alpha$ of path length $d_{r,l}$. To make sure in this backward deduction process the polarity of $\alpha$ corresponds to the same strategies as in the forward allocation process, we put $(d_{r,l})^\alpha$ in the denominator.

After the two stages, the amount of capacity deduction on each path $l$ for request $r$ $\Delta_{i,j}^{r,l}$ is determined, so the flow capacity $f_{i,j}^{r,l}$ is given by:
$$f_{i,j}^{r,l} = f_{max}^{r,l} - \Delta_{i,j}^{r,l}. \tag{S17}$$

$f_{i,j}^{r,l}$ is then the updated maximum flow on edges in the path, if it's smaller than the previous value. Note that not only the maximum flow values on following edges, but also on previous edges are updated. Essentially, the short-board constraint also narrows the flow on previous edges along the path, whose capacity allocation is then recalculated; since the flow on the current path is narrowed, recalculation may release extra capacity for other paths and hence the overall capacity allocation becomes more efficient.

Given that the path information is required in the algorithm, this method does not follow a predefined order in treating the edges, as opposed to the first method where all edges could be treated individually. Since lattice networks are regular and easy to be ordered, we used a fixed order in treating the edges and iterate with startovers, until no update of flow occurs on any edge, i.e. the headcount of no improvements equals the edge number. The iteration is guaranteed to converge if the (directed) paths are all non-cyclic; in practice we ensure that the two nodes in a request pair are not identical, thus the condition holds. The backward adaptive deduction process is summarized in Algorithm 3.

## 3 Discussion on scheme parameters

In this section, we discuss the parameters that are important in determining the performance of capacity allocation algorithms. The parameters are $X = \{l_{max}, k, \alpha, \beta\}$ for *proportional share* and *propagatory update*, and $X = k$ for *progressive filling*. We will focus on $\{l_{max}, \alpha, \beta\}$ here.

### 3.1 $l_{max}$ dependence

The maximal allowed number of paths on each edge $l_{max}$ characterizes the controllability of edge traffic. In Fig. S3 (a-b) we show the $k$ dependence of the total flow and average capacity utilization with different $l_{max}$. For fixed $k$, the total flow $F$ and average capacity utilization $U_{avg}$ show an increase as a function of $l_{max}$ at beginning. This performance gain can be attributed to the fact that each edge enables more paths passing through it. This increase then saturates at larger $l_{max}$ (for example, $l_{max} = 10$ here), in which scenario the edge removal balances the gain.

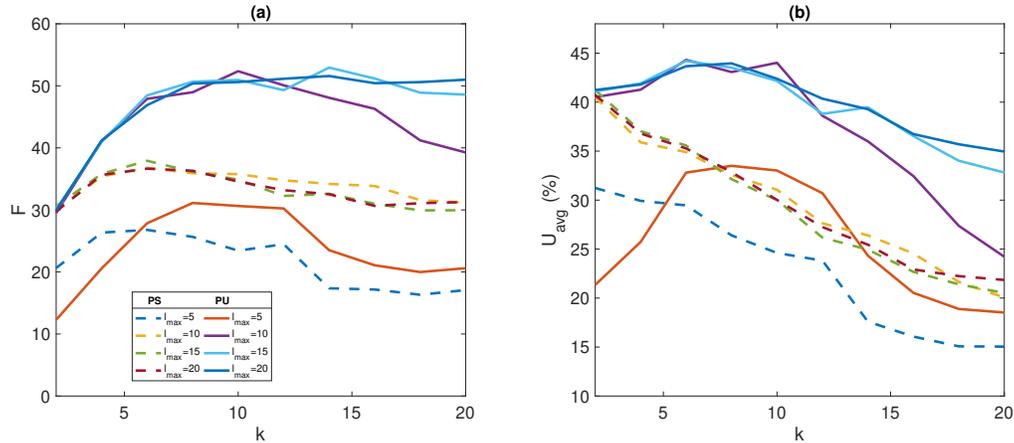

Figure S3: Routing performance for different $l_{max}$ and $k$. **(a).** Total flow $F$. **(b).** Average capacity utilization $U_{avg}$. The request distance is 3 and the other parameters are $\{\alpha, \beta\} = \{1, 0\}$ here. The PF algorithm performance is not shown here since it has no dependence on $l_{max}$.

### 3.2 $\alpha$ dependence

The parameter $\alpha$ determines the relation between capacity allocation and path length for the PS and PU algorithms, as we discussed in Section 2.1. Here we show the dependence of routing performance on $\alpha$. In the first scenario (in Fig. S4 (a-c)), when the request distance is large and $k$ is small, the utilized paths have identical length (shortest length) and the performance has no dependence on $\alpha$. While in the second case when the request distance is large and the paths spread out the network (in Fig. S4 (d-f)), the allocation process will prefer long paths at large $\alpha$ and the capacity of edges outside the shortest paths can be better utilized, leading to an increase of the $U_{avg}$. The path fairness $J_{path}$ seems to take the optimal value at $\alpha = 0$, in which case the capacity allocation will neither prefer short paths nor long paths.

### 3.3 $\beta$ dependence

The parameter $\beta$ determines the relation between capacity allocation and number of paths under different request for the PS algorithm, as we have discussed in Section 2.1. The throughput and traffic metrics (Fig. S5 (a-b))indicates that allocation on an edge without considering the number of paths for different requsts that going through that edge (*i.e.*, $\beta = 0$) is the best choice. However, the request fairness $J_{req}$ will be degraded in this case (Fig. S5 (c)).

## 4 Performance under network failures

In the main text, we have shown the total flow in the presence of network failures. Besides system throughput, the changes of other performance measures before and after network failures are shown below for the convenience of readers. Note that after network failures, some paths are not accessible, so some values of $\gamma$ are missing. In each figure, white/red bars indicate values before/after failures,

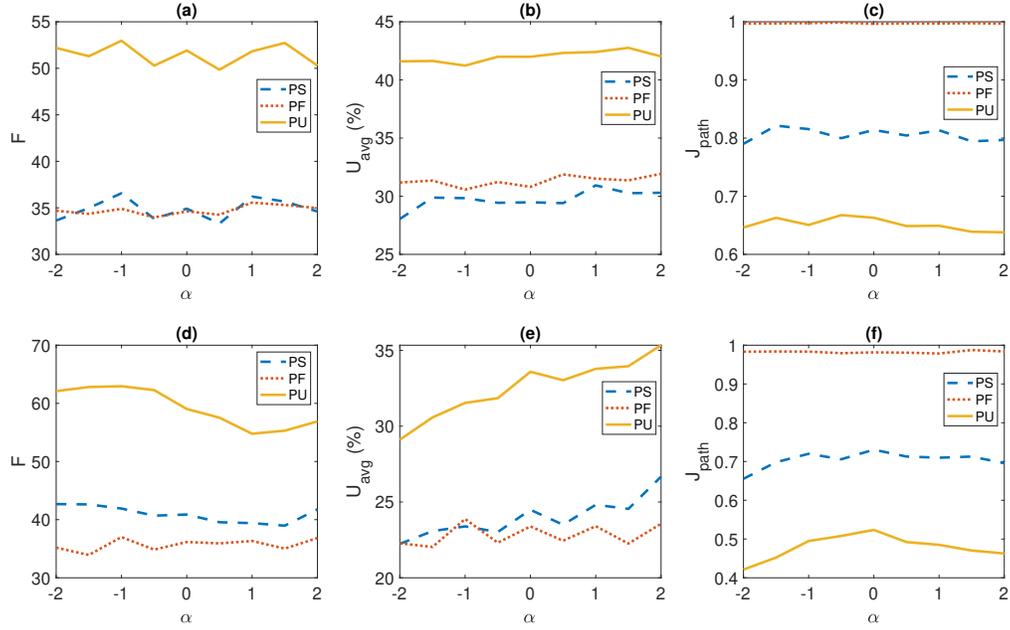

Figure S4: Routing performance with respect to $\alpha$. **(a-c).** Total flow $F$, average capacity utilization $U_{avg}$ and path fairness $J_{path}$ for request distance 2 and $k = 10$. **(d-f).** The metrics for request distance 3 and $k = 20$. $\{l_{max}, \beta\} = \{15, 0\}$ The PF algorithm performance is shown here for comparison.

respectively.

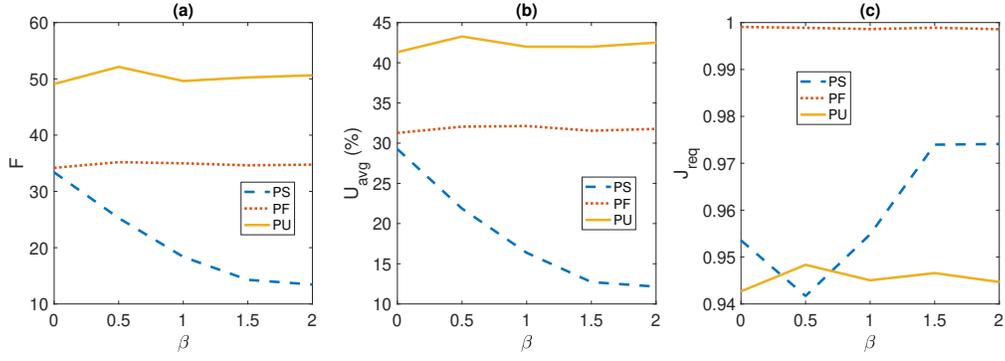

Figure S5: Routing performance with respect to $\beta$. **(a).** Total flow $F$. **(b).** Average capacity utilization $U_{avg}$. **(c).** Request fairness $J_{req}$. The request distance is 3 and the other parameters are $\{l_{max}, k, \alpha\} = \{15, 10, 1\}$ here.

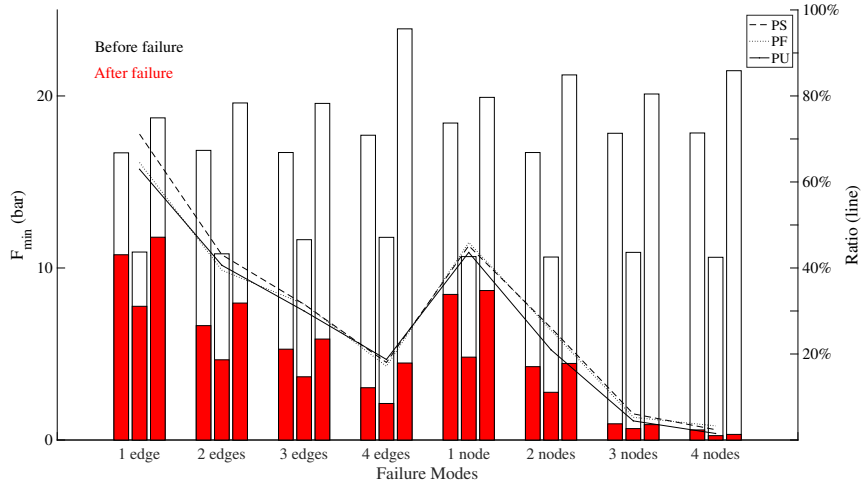

Figure S6: Robustness in the presence of network failures: minimum flow over requests $F_{min}$

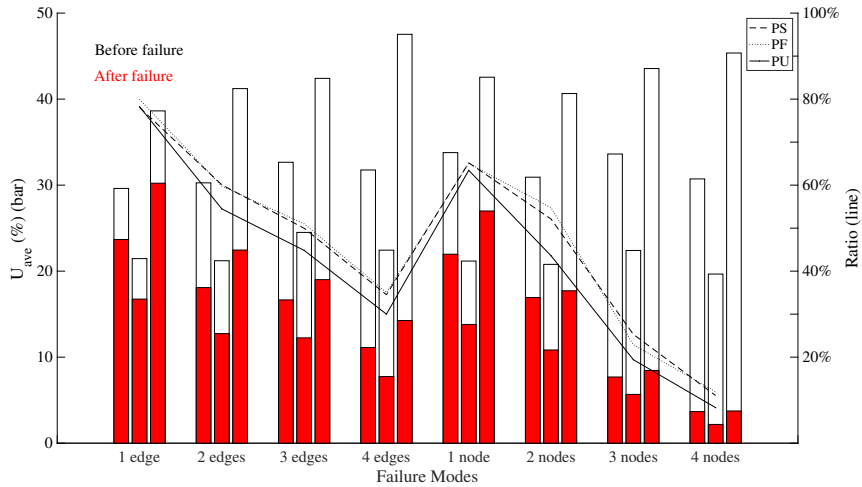

Figure S7: Robustness in the presence of network failures: average capacity utilization $U_{ave}$

11

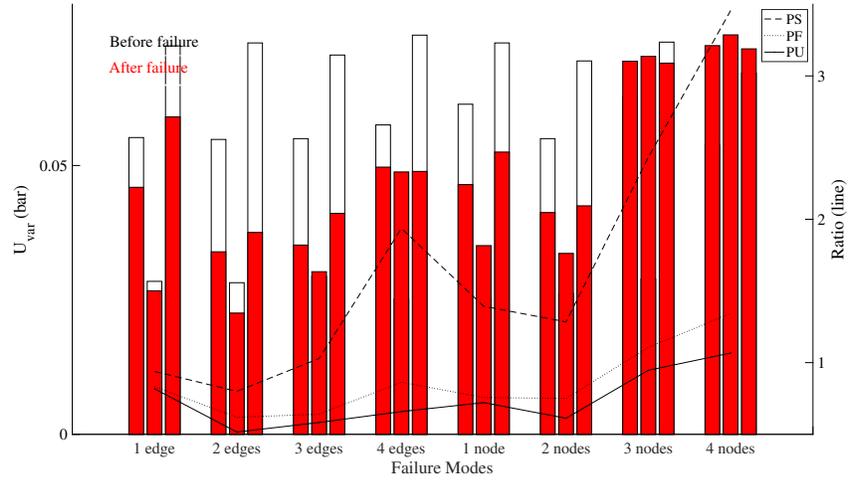

Figure S8: Robustness in the presence of network failures: variance of capacity utilization $U_{var}$

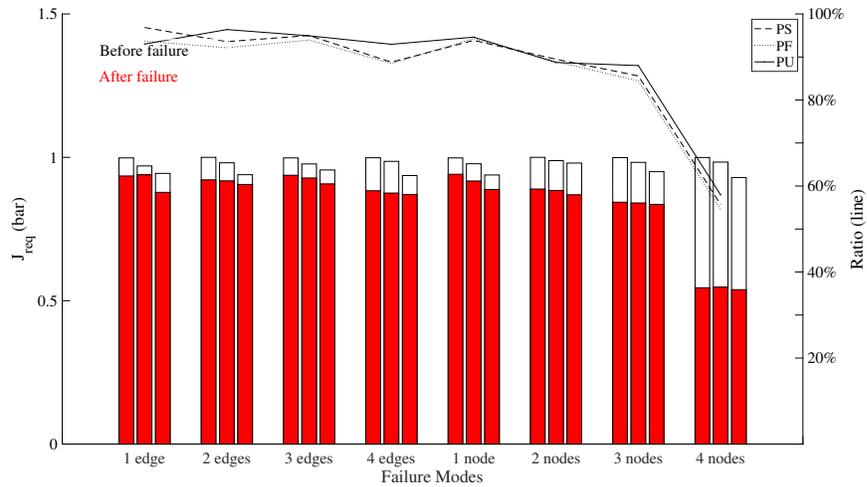

Figure S9: Robustness in the presence of network failures: fairness over requests $J_{req}$

12

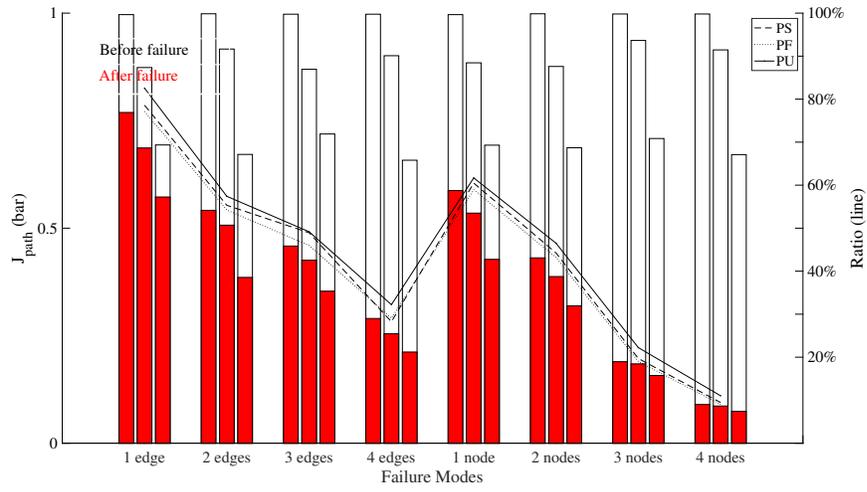

Figure S10: Robustness in the presence of network failures: fairness over paths $J_{path}$

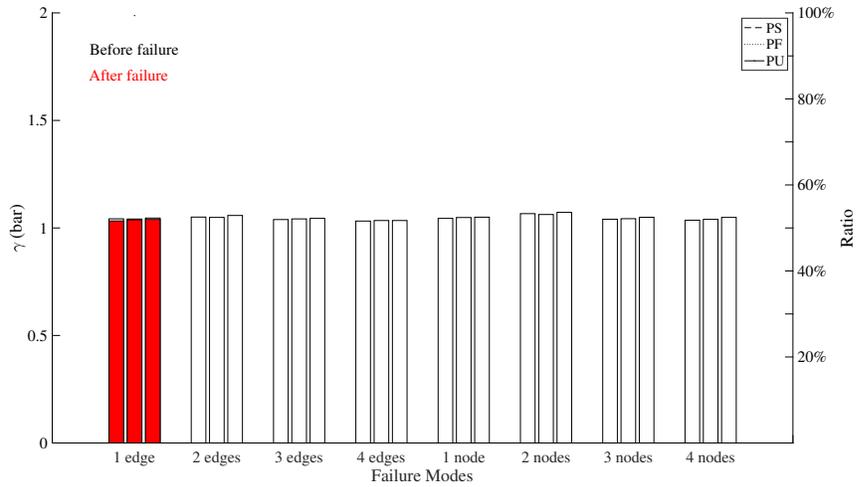

Figure S11: Robustness in the presence of network failures: $\gamma$

13



\end{document}